# Influence of Crystal Structure and Composition on Optical and Electronic Properties of Pyridinium-based Bismuth Iodide Complexes

Gisya Abdi,[a]* Marlena Gryl,[b]* Andrzej Sławek,[a] Ewelina Kowalewska,[a] Tomasz Mazur,[a] Agnieszka Podborska,[a] Krzysztof Mech,[a] Piotr Zawal,[a] Anurag Pritam,[a] Angelika Kmita,[a] Lulu Alluhaibi,[c] Alexey Maximenko,[c] Chakkooth Vijayakumar,[d] Konrad Szaciłowski[a]*

a. Academic Centre for Materials and Nanotechnology, AGH University of Krakow, Kawiory 30, 30-055 Kraków, Poland

b. Faculty of Chemistry, Jagiellonian University, Gronostajowa 2, 30-387 Kraków, Poland

c. National Synchrotron Radiation Centre SOLARIS, Czerwone Maki 98, 30-392 Kraków, Poland

d. Photosciences and Photonics Section, CSIR-National Institute for Interdisciplinary Science and Technology (CSIR-NIIST), Thiruvananthapuram 695 019, India.

*Corresponding authors: agisya@agh.edu.pl, marlena.gryl@uj.edu.pl, szacilow@agh.edu.pl

**Abstract**

This study investigates the impacts of structure and composition on the optical and electronic properties of a series of pyridinium-based bismuth iodide complexes. Organic substrates with various functional groups, such as 4-aminopyridine (4-Ampy), 4-methylpyridine (4-Mepy), 4-dimethyaminopyridine (4-Dmapy), and 4-pyridinecarbonitrile (4-CNpy) with different electron-donating and electron-withdrawing groups at the para position of the pyridine ring were employed. Crystallographic analysis reveals various bismuth iodide structures, including 1D chains and discrete 0D motifs. The optical band gap of these materials, identified via Diffuse reflectance spectroscopy (DRS) and verified with density functional theory (DFT) calculations, is influenced by the crystal packing and stabilising interactions. Through a comprehensive analysis, including Hirshfeld surface (HS) and void assessment, the study underscores the influence of noncovalent intermolecular interactions on crystal packing. Spectroscopic evaluations provide insights into electronic interactions, elucidating the role of electron donor and acceptor substituents within the lattice. Thermogravimetric differential thermal analysis (TG-DTA) indicates structural stability up to 250°C. Linear sweep voltammetry (LSV) reveals significant conductivity in the range of 10-20 mS/pixel at 298.15 K. X-ray absorption spectroscopy (XAS) at the Bi L3 edge indicates a similar oxidation state and electronic environment across all samples, underscoring the role of bismuth centres surrounded by iodides.

## Introduction

Lead halide perovskites have attracted significant attention as photovoltaic materials because of their remarkable photoconversion efficiency in solar cell devices. They also show promise for resistive switching devices, light-emitting devices, lasers, photodetectors, X-ray imaging, and thin-film transistors.[1] Corner, edge, or face-sharing of $MI_6$ octahedral fragments can give rise to zero-dimensional (0D), one-dimensional (1D), two-dimensional (2D) or even three-dimensional (3D) architectures.[2-4] The electronic and photoelectronic properties of these structures and their applications are heavily influenced by the counterion. However, concerns over the toxicity of lead halide complexes and the instability of lead (II) halide complexes, have propelled research into environmentally friendly alternatives.[5-11] Several perovskite-like compositions such as $A_3M_2I_9$, $A_3MX_6$, and $AMX_4$ (X = Cl, Br, I) have been proposed, where $Pb^{2+}$ is partially or fully replaced by metal cations such as $Bi^{3+}$/$In^{3+}$/$Sb^{3+}$/$Fe^{3+}$/$Mn^{2+}$/ $Sn^{2+}$/$Co^{2+}$/$Ge^{2+}$ and monovalent cations (A: organic methylammonium ($MA^+$) or inorganic cation e.g. $K^+$, $Rb^+$, $NH_4^+$, $K^+$, $Cs^+$, $Ag^+$, $Tl^+$).[12] For example, low-dimensional manganese-based perovskites were reported by Xiao et al. [13] which by phase transformation from 1D-$CsMnCl_3(H_2O)_2$ to 0D-$Cs_3MnCl_5$ result in shift from red to bright green luminescence emission. Furthermore, Anthony group[8] used these materials with ZnO in light harvesting and asserted that their photoluminescence is a morphology-dependent process. Band gap engineering by changing the size and content of perovskites is a popular research topic.[7, 14] The size of organic cation and its steric hindrance crucially impact the band gap values and consequently affect the electronic and photoelectronic properties of solid-state organometal structures and their practicable applications.[14] The light absorption peaks of the methylammonium lead iodide perovskite have been shown to shift when part of the organic cation is replaced or substituted with caesium or formamidinium and can enhance the photovoltaic performance in solar cells.[15] Peng et all. by changing dimensionality in the crystalline structure of caesium antimony halide perovskites ($Cs_3Sb_2I_9$) manipulated photoconversion efficiency in photovoltaic devices.[16]

In parallel, a wide range of bismuth(III) complexes has been reported, with several finding applications in photovoltaic and neuromorphic computing devices.[1, 17-21] Halobismuthate(III) anions of various organic salts form simple discrete mononuclear $[BiX_6]^{3-}$ structures, to more complicated binuclear $[Bi_2X_{10}]^{4-}$ and multinuclear anions such as $[Bi_4I_{16}]^{4-}$, $[Bi_4Br_{18}]^{6-}$, $[Bi_5Cl_{18}]^{3-}$, $[Bi_6I_{22}]^{4-}$, $[Bi_6Cl_{26}]^{8-}$, $[Bi_7I_{24}]^{3-}$ and $[Bi_8Cl_{28}]^{4-}$, and $[Bi_2X_9]^{3-}$ [22, 23]. The main factor that determines the structure and topology of the halobismutate(III) fragments is the weak interaction between bismuth-based building blocks and cations.

Here we present a detailed study on the subtle fine-tuning of pyridinium iodobismuthates(III) by modulation of these interactions through modification of the electronic structure of the cation while maintaining its geometric integrity as much as possible. For this study, we selected a series of structurally related pyridinium cations. Despite their almost identical geometry, these cations vary significantly in terms of their electronic structure, electron donor or acceptor characteristics, and dipole moment; nevertheless, they carry the same total charge. A comprehensive set of optical and electrical techniques, such as single-crystal XRD, UV-Vis absorption and reflectance, TG-DTA, XAS, linear sweep voltammetry (LSV), supplemented by theoretical methods included in DFT, Hirshfeld surface analysis (HS), and void analysis by CrystalExplorer, is available for characterisation of the aforementioned structures.

**Experimental section**

*Synthesis of pyridinium-based bismuth iodides*

The compounds were formed in the reaction described in Scheme 1. Initial concentrations of $BiI_3$, KI, and the appropriate pyridine derivative amounted to 0.66, 1, and 1 mmol, respectively. The reagents were dissolved in acetic acid/deionised water (1:1 ratio). The products were precipitated, then after 24 hours, the mixture was filtered, and the solid part was washed with the same solvent used in the reaction mixture.

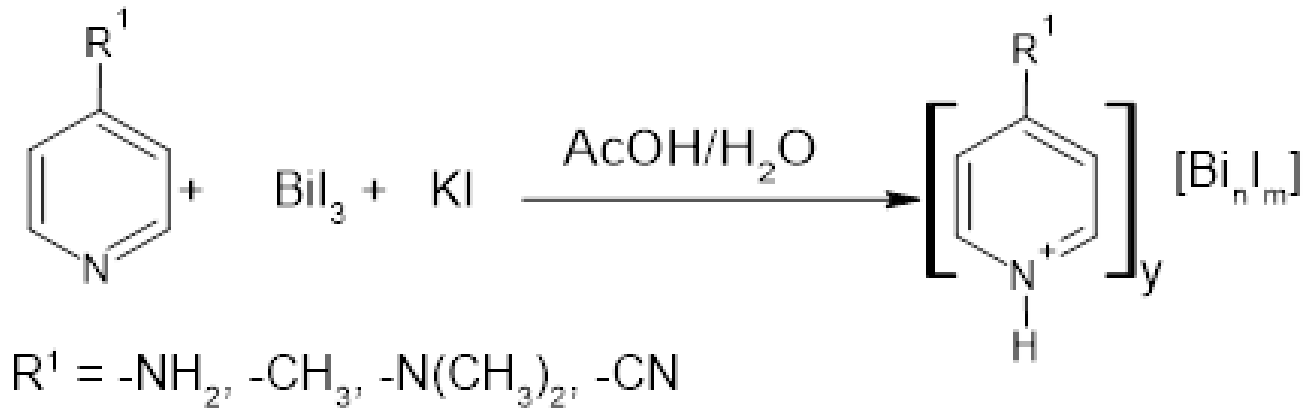

***Scheme 1**. Synthetic procedure and chemical reagents applied in the preparation of N-containing bismuth complexes.*

Recrystallisation was performed by dissolving the crystals in acetonitrile with n-hexane employed as an antisolvent. This mixture was left to evaporate over an extended period, allowing the crystals to grow. The solvothermal reaction of 4-cyanopyridine (1 mmol) and $BiI_3$ (0.66 mmol) in the presence of HI (1 mmol) at 60°C in methanol was used as an alternative method for the single crystal preparation of 4-cyanopyriidne, as it could not be achieved in acetonitrile. The elemental analysis (C:H:N) proved the presence of a methyl group in the structure of 4-CNpy$BiI_3$, which shows the methylation of the nitrogen atom in pyridine moiety rather than protonation (Table SI1). The resulting orange crystals were characterised by X-ray crystallography technique. Single-crystal measurements revealed that the cations comprised of an equal proportion of methylated pyridine and protonated entities.

*Characterisation techniques*

Using CASTEP software, the electronic structure was determined by first-principles density functional theory (DFT) using CASTEP software.[24, 25] A general gradient approximation (GGA) using a Pendrew-Burke-Ernzehof (PBE) functional was used to characterise non-local exchange correlation energies. CASTEP-specific (OTFG) pseudopotentials were exploited for computing.[26, 27] We included Van der Waals interactions using semi-empirical dispersion correction with a Tkatchenko-Scheffler (TS) scheme, while the spin-polarisation effects were omitted. Γ-point sampling of the irreducible part of the first Brillouin zone was used. Geometry optimisation was performed prior to band structure calculations. For geometry optimisation and electron spectroscopy, the electronic minimisation parameter of the total energy per atom was set to $10^{-5}$ and $10^{-6}$ eV, respectively. Gaussian-like Fermi smearing was used. The distance cutoff point for bond populations was set to 3 Å. Reflectance measurements of synthesised structures were performed by the LAMBDA 750 UV/vis/NIR spectrophotometer (PerkinElmer Inc.). A single beam UV-vis spectrophotometer Agilent 8453 was applied to measure the absorption of samples in different solvents. The I-V plots were recorded using a BioLogic SP-300 potentiostat. Single-crystal XRD experiments were performed on XtaLAB Synergy-S X-ray diffractometer using MoKα radiation at 100-130 K. Data collection and reduction were performed with CrystalisPro 1.171.41.122a (Rigaku OD, 2021). First, the structures were solved with direct methods using SHELXT,[28, 29] then, refinement was carried out with SHELXL[29] as incorporated into a WinGX package.[30]

For *ex situ* X-ray absorption spectroscopy (XAS) measurements, the powder samples were ground with mortar and pestle, dispersed in microcrystalline cellulose, and tightly packed between two Kapton tapes. The spectra at the Bi L3-edge were collected at the bending magnet ASTRA beamline of the National Synchrotron Radiation Centre SOLARIS.[31] The measurements were made in transmission mode using an incident photon beam provided by a modified Lemonnier type double crystal monochromator equipped with a Si(400) crystal pair as monochromators. Three consecutive scans for each sample were processed and analysed using the ATHENA program from the DEMETER software package.[32] The energy calibration and alignment of the collected spectra were made on the basis of Pb foil reference spectra, recorded between sample scans. The experimental absorption edge was determined on the basis of the position of the maximum of the first derivative of the spectrum, which is equivalent to the intersection with 0 of the second derivative. XAS spectra were calculated using FDMNES software,[33] using the density functional theory with the local spin density approximation (LSDA) method. Bi $L_3$-edges were calculated with Finite Difference Method for XAFS (FDMX)[34] using dipole ($\Delta l = \pm 1$) and quadrupole transitions ($\Delta l = 0, \pm 2$) applying multi-electronic time-dependent DFT (TDDFT) extension with a local kernel. The cluster radius was set to 6 Å. Spin-orbit coupling effects were also included, whereas relativistic effects were omitted. Lorentzian convoluted spectra have been presented.

***Table 1**. Selected electronic and topological parameters of the cations under study.*

| cation | $\mu^a$ / D | $q_N^b$ | $S^c$ / nm$^2$ | $V^d$ / nm$^3$ | $O^e$ | $\kappa^e$ |
|---|---|---|---|---|---|---|
| Pyridinium | 1.88 | -0.123 | 97.62 | 69.14 | 1.198 | 4.166 |
| Methylpyridinium | 2.85 | -0.123 | 116.47 | 85.94 | 1.237 | 5.143 |
| Aminopyridinium | 0.50 | -0.150 | 110.50 | 80.25 | 1.228 | 5.143 |
| Dimethylaminopyridinium | 4.89 | -0.147 | 144.02 | 112.93 | 1.276 | 7.111 |
| Cyanopyridinium | 8.32 | -0.107 | 114.21 | 82.39 | 1.247 | 6.125 |
| N-methylcyanopyridinium | 7.65 | -0.039 | 132.68 | 99.06 | 1.281 | 7.111 |

*[a] dipole moment; [b] Mulliken charge at aromatic nitrogen; [c] Connolly's surface area; [d] Connolly's molecular volume; [e] ovality; [f] shape factor.*

TG-MS tests were performed using a mass spectrometer (Hiden Analytical) integrated with the thermogravimetric analyzer SDT Q600 of TA Instruments, under an inert atmosphere (He 99.9999) with a constant heating rate of 5°C·min$^{-1}$ from ambient temperature to 500°C. Volatile materials resulting from the decomposition of compounds were analysed using mass spectrometry. The operating conditions of the ion source of the MS analyser were as follows: electron energy 70 eV, emission current 400 μA. The evolution curves of gasses release are tracked by the selected ion recording (MID) mode. Thin layers were prepared on ITO/glass surfaces purchased from Ossilla Ltd. and the Polos SPIN150i system was used to spin-coat materials on the surface. Leica EM ACE600 high-vacuum sputter coating was applied for sputtering metals as the top electrode on sample layers on ITO/glass. The thickness of the layer was measured using the Bruker DektakXT needle profiler.

*Device fabrication*

Thin layer devices were prepared using the spin coating technique. ITO/glass surfaces were washed in an ultrasonic bath using deionised water and 1 weight percent detergent (Hellmanex), then intensively rinsed with DI water and isopropyl alcohol, dried with $N_2$, and cleaned using $O_2$ plasma for 15 minutes. Stock solutions were produced by dissolving 150 mg of each pyridinium-based bismuth iodide compound in 0.5 ml of DMF. The clean and dry ITO substrates were spin coated with synthetic materials and then the layer on the substrate was immediately transferred to a heater and annealed at 100°C for 15 min. The fabrication of the metal-semiconductor contacts was performed by sputtering copper contacts with the desired nanometric sizeon material covered by a shadow mask with an electrode area of 1 $mm^2$. The concentration of the complex in the solvent and the rotation rate of the spin coater were optimized to provide conditions that allowed the preparation of complex layers of an average thickness of 150±25 nm (measured using contact profilometry).

**Results and Discussion**

*Cation Selection and Characterisation*

Both $Pb^{2+}$ and $Bi^{3+}$ have weakly bound 6s electrons, resulting in high polarisability that facilitates distortion and aggregation of $MI_x^{n-}$ polyhedra. This property gives rise to a wide array of possible geometric structures and also modulates band gap energy, the mobility of the charge carrier and other electronic properties of the corresponding materials.[35] In this study, we explore the subtle influence of counterions on the structure and properties of iodobismuthates. We selected a series of closely related pyridinium cations: pyridinium, 4-methylpyridinium, 4-aminopyridinium, 4-dimethylaminopyridinium, and 4-cyanopyridinium. Besides pyridinium, all these cations share a similar geometry, but display distinctive electronic structure. Preliminary characterisation of the cations was performed using DFT calculations and topological analysis of the resulting geometries. This included measurements of Connolly's molecular area and volume, ovality of the molecule, and other shape attributes.

As expected, the function groups at position 4 (*para*), despite their structural similarity, significantly influence the distribution of electric charge within each cation (Figure 1). This influence directly impacts the dipole moments of each cation. On the other hand, the point charges at the nitrogen remain virtually identical across the cations (Table 1).

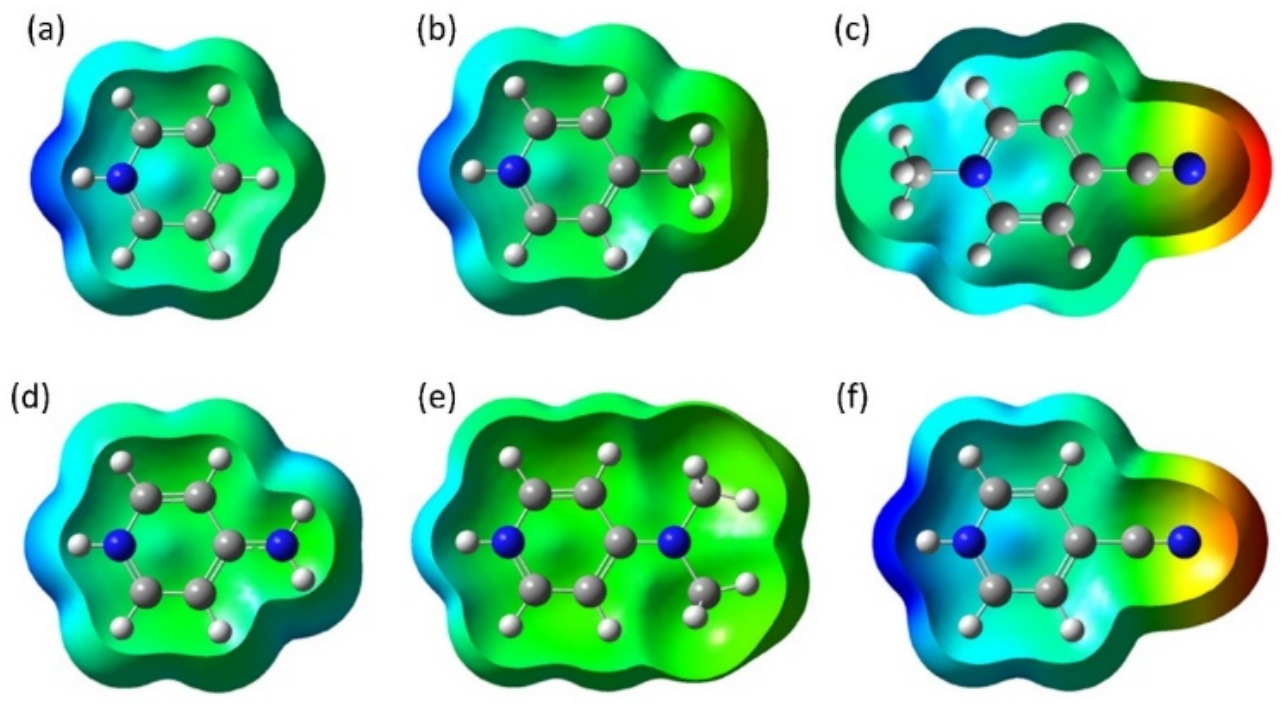

***Figure 1**. Converged molecular geometries of the studied cations and electrostatic potential distribution mapped onto electron density isosurfaces: a) pyridinium, b) methylpyridinium, c) N-methylcyanopyridinium, d) aminopyridinium, e) dimethylaminopyridinium, and f) cyanopyridinium.*

***Table 2.** Crystal data and refinement details.*

| Compound | **4-AmpyBiI$_3$ (1)** | **4-MepyBiI$_3$ (2)** | **4-DmapyBiI$_3$ (3)** | **4-CNpyBiI$_3$ (4)** |
|---|---|---|---|---|
| Chemical formula | $C_5H_7BiI_4N_2$ | $C_{18}H_{24}Bi_2I_9N_3$ | $C_{28}H_{44}Bi_2I_{10}N_8$ | $C_{14}H_{16}Bi_2I_8N_4O$ |
| $M_r$ | 405.85 | 1842.46 | 2179.67 | 1689.47 |
| Crystal system, space group | orthorhombic, *P*bcn | orthorhombic, $P2_12_12_1$ | orthorhombic, *P*nma | monoclinic $P2_1/n$ |
| Lattice parameters (Å)/° | a = 12.2049(2)<br>b = 14.8651(3)<br>c = 7.6069(1) | a = 9.2443(1)<br>b = 10.6401(1)<br>c = 37.9400(3) | a = 17.9971(6)<br>b = 20.3282(7)<br>c = 13.9362(4) | a = 14.9874(3)<br>b = 11.4555(2)<br>c = 20.4765(5)<br>β = 107.648(2) |
| T (K) | 100(2) | 130(2) K | 130(2) | 298(2) |
| V (Å$^3$) | 1380.10(4) | 3731.79(6) | 5098.5(3) | 3350.1(1) |
| Z | 4 | 4 | 4 | 4 |
| $D_x$(g/cm$^3$) | 3.902 | 3.279 | 2.840 | 3.350 |
| Radiation type | MoKα | MoKα | MoKα | MoKα |
| μ (mm$^{-1}$) | 21.686 | 16.877 | 12.977 | 17.877 |
| Theta range | 2.740° to 32.874° | 2.147° to 33.429° | 2.102° to 33.495° | 2.320 to 33.486° |
| Diffractometer | Rigaku Synergy S | Rigaku Synergy S | Rigaku Synergy S | Rigaku Synergy S |
| Absorption correction | multi-scan | multi-scan | multi-scan | multi-scan |
| Crystal size (mm) | 0.10x0.08x0.03 | 0.28x 0.12x 0.06 | 0.20x0.11x0.03 | 0.20x0.16x0.08 |
| Data/restraints/parameters | 2472/0/57 | 13858/0/292 | 9690/0/234 | 11871/0/265 |
| R(int) | 0.1002 | 0.1009 | 0.0677 | 0.0604 |
| Goodness-of-fit | 1.081 | 0.940 | 1.013 | 1.037 |
| Final R indices [I>2sigma(I)] | $R_1$ = 0.0202, $wR_2$ = 0.0431 | $R_1$ = 0.0405, $wR_2$ = 0.0951 | $R_1$ = 0.0269, $wR_2$ = 0.0576 | $R_1$ = 0.0405, $wR_2$ = 0.0892 |
| $\Delta\rho_{min}/\Delta\rho_{max}$ (e/Å$^3$) | -1.169/1.179 | -3.495/3.440 | -1.772/2.147 | 1.772/-1.325 |
| Absolute structure parameter | - | -0.022(2) | - | - |

Despite significant variations in electrostatic potential distributions and the resulting differences in dipole moments, it is noteworthy that the point charges at the aromatic nitrogen atoms remain quite similar. These charges adjust on the nature of the substituent at the para position. For example, electron-donating groups, such as amino and dimethylamino groups, lead to a slight increase in the point charge of the nitrogen atom. Conversely, the presence of electron-withdrawing cyano substituents results in a minor decrease in the charge.

Topological analysis allows for the observation of intuitive changes in molecular surfaces and volumes. However, the changes in these charges are not substantial, with the exception of the smallest pyridinium cation. However, this cation is excluded from further analysis as a result of the instability of its product. Importantly, all cations exhibit a rather similar shape, as indicated by the ovality factor *O.* This factor (defined in eq. 1), relates ovality to molecular volume, *V* and molecular surface, *S.* It is scaled in such a way that the ovality of a perfect sphere is equal to unity:

$$O = \frac{S}{4\pi\left(3V/4\pi\right)^{\frac{2}{3}}} \qquad (1)$$

In addition, all the compounds studied have similar values of the shape factor *κ*, which gauges the degree of molecular branching. Therefore, it is reasonable to infer that the observed variations in the structure and properties of pyridinium iodobismuthates result mainly from electronic or electrostatic interactions within the lattice with only minor contribution from steric or geometric factors.

*Synthesis and Stability*

Bismuth iodide readily precipitates salts with organic cations in the presence of an excess of iodide ligands. This reaction, historically known as the Dragendorff reaction, was originally used for the detection and identification of alkaloids.[36] It now serves as a key synthetic protocol for creating bismuth-based materials.[35] A series of pyridine-based structures were selected (Scheme 1) and their reaction with bismuth iodide under acidic conditions was examined. As a result of the low solubility of the products in water, they can be easily separated with relatively high yields. It is worth mentioning that the same reactions with pyridine were examined but the resulting product showed low stability and was always polycrystalline. These instability issues also extend to elevated temperatures, rendering these two compounds unsuitable for further device-oriented study. Interestingly, the reaction with 4-cyanopyridine in methanol produced an unexpected product: a compound containing *N*-methyl-4-cyanopyridinium. The unexpected methylation of the pyridine was likely due to the strong Lewis acid character of bismuth iodide.

The results of thermogravimetry-differential thermogravimetry (TG-DTG) experiments are shown in Figure SI1. Pristine pyridinium iodobismuthate ($pyBiI_3$) starts to decompose at 93.3°C (Figure SI1). However, the resulting products of the functionalised pyridine in Scheme 1 show higher stability, as depicted in Figure SI1 and summarised results in Table SI2. The DTG curves became steeper above 230°C, indicating the decomposition and evaporation of the components of the structures up to ~ 400°C. Among the bismuth complexes, 4-CNpyBiI$_3$ has the lowest stability with two-steps mass loss. Decomposition began after 200°C with a mass loss of 23 % occurring between 24 - 246 °C and 77 % between 246 - 400°C (Figure SI1d). On the contrary, the decomposition of the other three samples starts after 230°C and they showed a one-step decomposition. In most cases, the observed weight loss is attributed to dehydration followed by sublimation of the compound, as the mass loss shows zero and no residue left after decomposition at the end of experiments. Thermogravimetric analysis / mass spectrometry (TGA-MS) of 4-DmapyBiI$_3$ (Figure SI2a) shows signals related to $CH_2I^+$ (143), $I^+$(127), $CH_3I^+$(142), $CH_3^+$(15), in other samples the gas analysis did not show any strong signals in the mass spectra, which probably is due to sublimation of the samples and not complete decomposition at the given temperature. GC-MS analyses of the headspace gas indicated the presence of volatile organic compounds, including methyl iodide,, iodide and pyridine just in the case of 4-MepyBiI$_3$ (Figure SI2b).

***Table 3***. *Bi-I bond lengths in bismuth iodide-based structures.*

| **4-AmpyBiI$_3$** | [Å] | **4-MepyBiI$_3$** | [Å] | **4-CNpyBiI$_3$** | [Å] |
|---|---|---|---|---|---|
| Bi1-I2 | 3.196 | Bi1-I1 | 2.929 | Bi1-I1 | 2.881 |
| Bi1-I3 | 2.962 | Bi1-I2 | 2.937 | Bi1-I2 | 2.900 |
| **4-DmapyBiI$_3$** | [Å] | Bi1-I3 | 2.984 | Bi1-I3 | 3.354 |
| Bi1-I1 | 2.933 | Bi1-I4 | 3.241 | Bi1-I4 | 3.129 |
| Bi1-I2 | 2.910 | Bi1-I5 | 3.299 | Bi2-I5 | 3.322 |
| Bi1-I3 | 3.232 | Bi1-I6 | 3.254 | Bi2-I6 | 2.936 |
| Bi1-I4 | 3.330 | Bi2-I4 | 3.203 | Bi2-I7 | 2.941 |
| Bi1-I8 | 3.068 | Bi2-I5 | 3.298 | Bi2-I8 | 2.921 |
| Bi2-I3 | 3.305 | Bi2-I6 | 3.271 | Bi1-I5#1 | 3.030 |
| Bi2-I4 | 3.278 | Bi2-I7 | 2.921 | Bi1-I3#1 | 3.370 |
| Bi2-I5 | 2.929 | Bi2-I8 | 2.952 | | |
| Bi2-I6 | 3.051 | Bi2-I9 | 2.989 | | |
| Bi2-I7 | 2.906 | | | #1 –x+1, -y+1, -z+1 | |

*Crystal Structure Determination and Analysis*

The single-crystal XRD technique was employed to examine the materials structural features of the obtained (Table 2). The crystal phases of 4-AmpyBiI$_3$, 4-MepyBiI$_3$, 4-DmapyBiI$_3$ all belong to the orthorhombic crystal system. Of these, three are centrosymmetric and one is chiral. 4-CNpyBiI$_3$ (4) crystallises in a monoclinic P21/n space group. The four crystal structures containing different bismuth-iodide fragments are presented in Figure SI3, Figure SI4.

4-AmpyBiI$_3$ is based on infinite $[BiI_4]_n^{n-}$ chains where edges are shared, and separated by stacked 4-aminopyridinium cations. These cations are arranged in an antiparallel fashion with an average separation of ca. 3.5 Å between individual rings, a characteristic value for systems with moderate interactions.[37] This arrangement forms pseudo-herringbone assemblies. Notably, the structure is formed primarily by hydrogen bonds and iodide bridges, which determine the packing of the structural components. The cations form three different sets of hydrogen bonds with iodide anions of the $[BiI_4]_n^{n-}$ chain: (i) bonds of 3.06 Å between the hydrogen atom attached to the N8 atom in the aromatic ring, (ii) 2.97 Å bond that involve hydrogen atoms in the amino groups, and (iii) a complex set of close contacts (H···A distances in the range 3.1-3.3 Å) involving hydrogen atoms at the aromatic ring.

The asymmetric unit of the 4-AmpyBiI$_3$ structure is illustrated in Figure 2a. The inorganic part of the crystal structure is formed by the bismuth iodide fragments of $[BiI_4]^-$, interconnected by the edges of the octahedra in a one-dimensional fashion (Figure 3a). Centrosymmetric crystal structure belongs to the Pbcn orthorhombic space group. The I2 iodide bridges two Bi centres (3.196 Å) while I3 are terminal iodide ligands that form the shortest Bi-I bonds (equal to ~ 2.962 Å) (as detailed in Table 3). The amino group in the para position of the pyridinium cation forms two hydrogen bonds. The bond with terminal iodide (N4-H4B···I3) is shorter ( 2.97 Å) compared to the bond between protonated nitrogen in the pyridinium cation and terminal hydrogen (N8-H8···I3), which measures 3.06 Å. When analysing the geometry of weak interactions, it was revealed that the C7-H7···I2 is the weakest interaction with the smallest <DHA angle (around 114.5 degrees), and the longest distance between hydrogen (H7) and I2, ~ 3.31 Å (Table SI3).

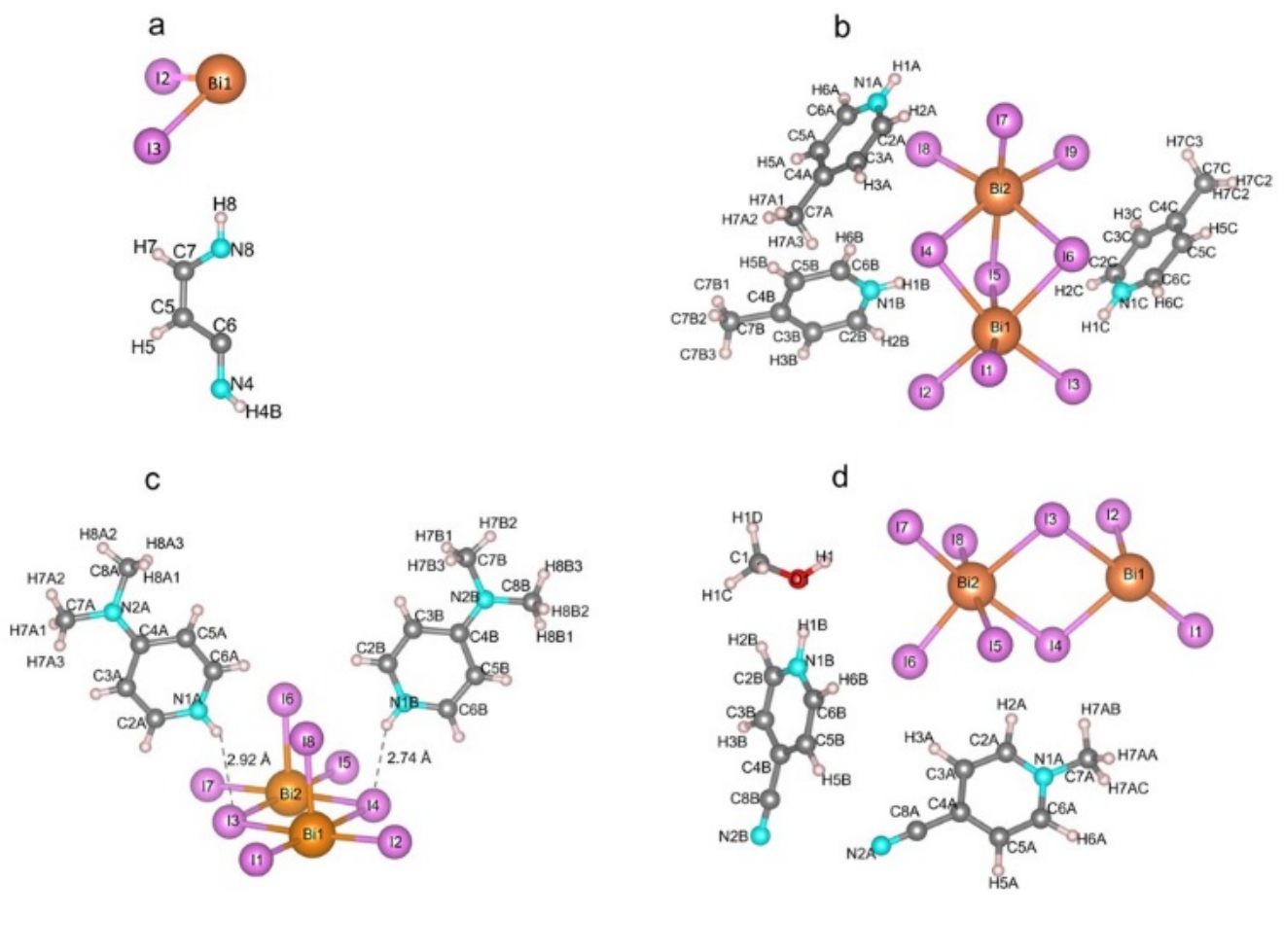

***Figure 2**. Illustration of the asymmetric unit composed of (a) 4-AmpyBiI$_3$ , (b) 4-MepyBiI$_3$, (c) 4-DmapyBiI$_3$, and (d) 4-CNpyBiI$_3$.*

The asymmetric unit of the 4-MepyBiI$_3$ structure, shown in Figure 2b, includes geometric features and hydrogen-bonding interactions with three 4-methylaminopyridinium and a 0D bismuth iodide fragment, $[Bi_2I_9]^{3-}$. The resulting crystal phase is noncentrosymmetric with crystal structure belonging to the Sohncke P212121 space group. According to a survey of the Cambridge structural database (CSD) by Teruo Matsuura, the probability of the formation of chiral crystalline structures from achiral molecules is only around 8 %.[38] In this structure, the Bi1-I1, Bi1-I2, Bi2-I7 and Bi2-I8 bonds are in equatorial positions with bond lengths less than 2.95 Å, which are shorter than Bi-I3 and Bi-I9 which occupy the axial positions (~ 2.98 Å). The Bi2-I4 bond, which serves as a bridge-axial bond, is distorted and shorter than the Bi1-I5, Bi2-I5 and Bi-I6 bonds, which serve as halide bridges between two Bi centres (Table 3). Three crystallographically independent classical hydrogen bonds are detectable in the structure: N1A-H1A···I9 (~ 2.7 Å), N1B-H1B···I5 (~ 2.74 Å), and N1C-H1C···I3 (~ 2.72 Å), respectively (see Table SI6). These interactions and other weak C-H···A interactions result in a packed crystal structure as shown by void analysis. The perspective view of the unit cell content is shown in the ac-plane in Figure 3e.

Figure 2c illustrates the asymmetric unit of the 4-DmapyBiI$_3$ structure, highlighting the geometric parameters and hydrogen bonding interactions between two 4-dimethylaminopyridinium cations and a bismuth iodide fragment $[Bi_2I_{10}]^{4-}$ formed through edge sharing of two octahedral units. Bi-I bonds, involving I2, I1, I5 and I7, in equatorial sites are the shortest (~ 2.9 ), whereas Bi-I4 and Bi-I3 bonds , where I3 and I4 act as bridges between two Bi centres, are the longest (~3.23 Å). The Bi-I bonds with the iodide in axial positions (I8 and I6) are of moderate length (around 3.05 Å), as seen in Table 3. Notably, the length of the C4A-N2A bond is 1.39 Å, which is close to the N1A-C2A bond length (~1.4Å). This indicates the double bond character of this connecting bond and the possible resonance of the dimethylamino group with the aromatic systems. This results in partial delocalization of charge density through sp$^2$ hybridisation. This conclusion is qualitatively consistent with the DTF data (Figure 1), which shows a lower positive charge at the pyridinium moiety, compared to other studied systems.

The bridging I3 and I4 atoms participate in the strongest hydrogen bonds found in this system, formed with a N1 atom in the pyridinium cations (A and B) acting as the donor (D). Table SI4 shows the geometry of weak interactions, distance between donor-hydrogen (D-H), hydrogen···acceptor (H···A), donor···acceptor (D···A) and donor-hydrogen-acceptor angle (<DHA). All the remaining interactions in this structure utilise carbon atom as hydrogen bond donor; thus they can be classified as weak in strength. Figure 3d illustrates the unit cell framework of 4-DmapyBiI$_3$, which shows ionic and cationic moieties distributed in a 3D structure, visualised by VESTA (Visualisation for Electronic Structural Analysis).[39]

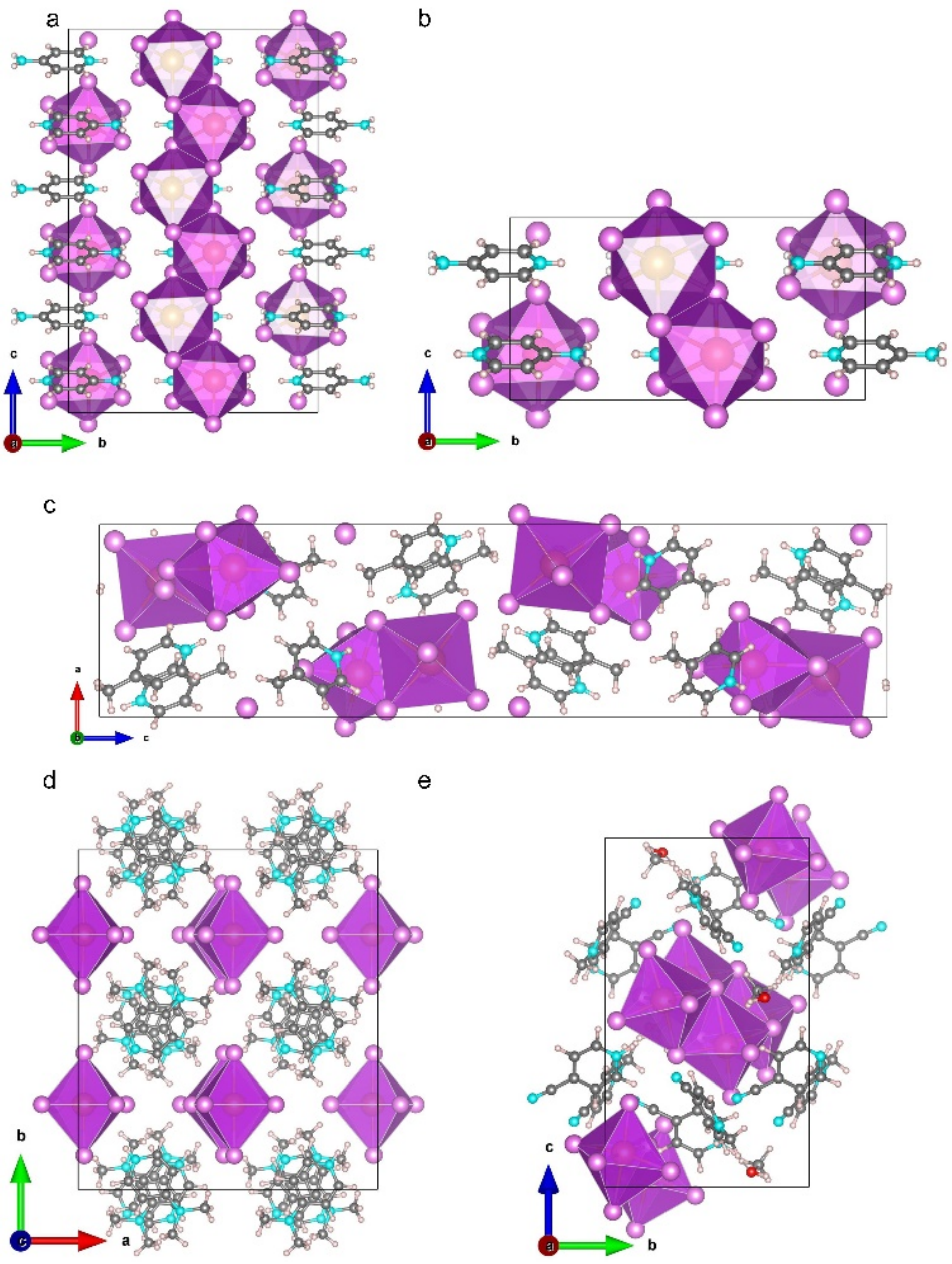

***Figure 3***. *(a) 4-AmpyBiI$_3$ side view of 1D growth of bismuth iodide fragments in crystal structures. Perspective view of the unit cell (b) infinite BiI$_6$ octahedra in 4-AmpyBiI$_3$; (c) by face sharing of two BiI$_6$ octahedra in 4-MepyBiI$_3$, and by edge sharing of (d) two BiI$_6$ octahedra in 4-DmapyBiI$_3$ (e) four BiI$_6$ octahedra in 4-CNpyBiI$_3$.*

***Table 4**. Physiochemical parameters calculated by void analysis for crystalline derivatives of pyridinium-based structures.*

| Entry | Unit cell volume (Å$^3$) | Void Volume (Å$^3$) | Void percentage | Area (Å$^2$) | Globularity (G) | Asphericity (Ω) |
|---|---|---|---|---|---|---|
| 4-AmpyBiI$_3$ | 1380.097501 | 59.66 | 4.3 | 240.29 | 0.307 | 0.07 |
| 4-MepyBiI$_3$ | 3731.788541 | 344.51 | 9.23 | 1198.16 | 0.198 | 0.658 |
| 4-DmapyBiI$_3$ | 5098.540065 | 509.14 | 9.98 | 1603.54 | 0.192 | 0.035 |
| 4-CNpyBiI$_3$ | 3350.1191 | 361.68 | 10.79 | 1119.87 | 0.219 | 0.156 |

The 4-CNpyBiI$_3$ complex displays a different crystalline structure compared to its pyridinium-based analogues. Crystal growth was only possible in solvothermal reaction of the starting materials in methanol as the solvent, resulting in appearance of 4-cyano-N-methylpyridinium along with 4-cyanopyridinium moieties in the crystal structure. Figure 2d shows the asymmetric unit comprising the $[Bi_4I_{16}]^{4-}$ ion, the pyridinium cation, and *N*-methyl pyridinium cation. This results in a centrosymmetric crystalline structure of 4-CNpyBiI$_3$ in a monoclinic system, as shown in Figure 3e. Weak interactions, including hydrogen bonds are listed in Table SI6. The distance between Bi1-Bi2 is 4.73 Å, similar to its electron-donating functional group analogues. In Figure 2d, Bi1-I3(3.354 Å) and Bi1-I4 (3.129Å) occupy the equatorial sites and bridge Bi units. The Bi2-I5 oriented at the axial site shows a distortion due to the bridging role in the structure that leads to 3.32 Å Bi-I bond length. Bi1-I2 and Bi2-I8 are in axial positions with bond lengths 2.9 Å, and 2.92 Å, respectively. On the contrary, Bi1-I1, Bi2-I6, Bi2-I7 at the equatorial positions show bond lengths of 2.88 Å, 2.936 Å, and 2.941 Å respectively. The latter two bond lengths do not align with expectations for equatorial sites. A possible explanation for this behaviour is the occupation of the media around Bi2-I6 and Bi-I7 by the methanol molecule and the organic moieties are located in close proximity to Bi2-I6 and Bi2-I7, which act as acceptors by sharing lone pairs, resulting in longer bond lengths with the centre Bi centre (~ 4 Å), as summarised in Table SI6.

The iodobismuthate moiety within the 4-CNpyBiI$_3$ structure comprises four Bi-I octahedra that share their edges, forming isolated $[Bi_4I_{16}]^{4-}$ units, which are surrounded by two types of pyridinium cations. Analysis of the geometry of the interactions reveal a weak interaction between protons in the methyl group attached to nitrogen pyridinium nitrogen and bridging Iodide (C7A-H7AB···I4 ) with bond length 3.17 Å. Additionally, there is a moderate hydrogen bond between the hydrogen atom in the pyridinium moiety and the hydroxyl group in methanol (N1B-H1B···O1); however, no significant interaction with the iodide groups in the $[Bi_4I_{16}]^{4-}$ ion was detected. The OH group in methanol exhibits a shorter spatial distance with I7 in the equatorial position (2.91 Å). The proton in the ortho position of *N*-methylpyridinium (C2A-H2A) displays weak interactions with I4 and I5, located at the bridging positions (Table SI6). On the contrary, protons in species B (C2B-H2B and C3B-H3B) primarily interact with iodides in equatorial sites (I6 and I7) at terminal positions. Angle measurements revealed that C3B-H3B···I7, C2A-H2A···I4, C7A-H7AB···I4, C6A-H6A···I2 with <DHA more than 150°.

The observed results reveal that, depending on the dipole moment of the pyridinium cation, various modifications of the structures were obtained: 1D polymeric chains based on edge-sharing two $[BiI_6]^{3-}$ octahedra in the case of 4-Ampy and various 0D coordination clusters resulting from different fusion patterns (edge- and face-sharing) of $[BiI_6]^{3-}$ units and ultimately building complex 3D ionic lattices with pyridinium counterions.

Close contacts and other noncovalent interactions were studied using Hirshfeld analysis. This analysis involved mapping the surface the normalized contact distance ($d_{norm}$) computed based on the inner ($d_i$) and outer ($d_e$) distances of the Hirshfeld surface (HS) to the nearest nucleus (displayed in Figure 4). This method allowed an examination of the non-covalent interactions existing in the synthesised structures listed in Table SI 7, using CrystalExplorer 17.[40] The surfaces have been mapped with $d_{norm}$, a shape index ranging from -1.000 to 1.000 Å, and curvedness ranging from -4.000 to 0.4000 Å (Figure SI5). Intermolecular interactions are differentiated through colour coding. The red, white, and blue spots represent intermolecular interactions with distances that are less than, equal to, and greater than van der Waals lengths, respectively.

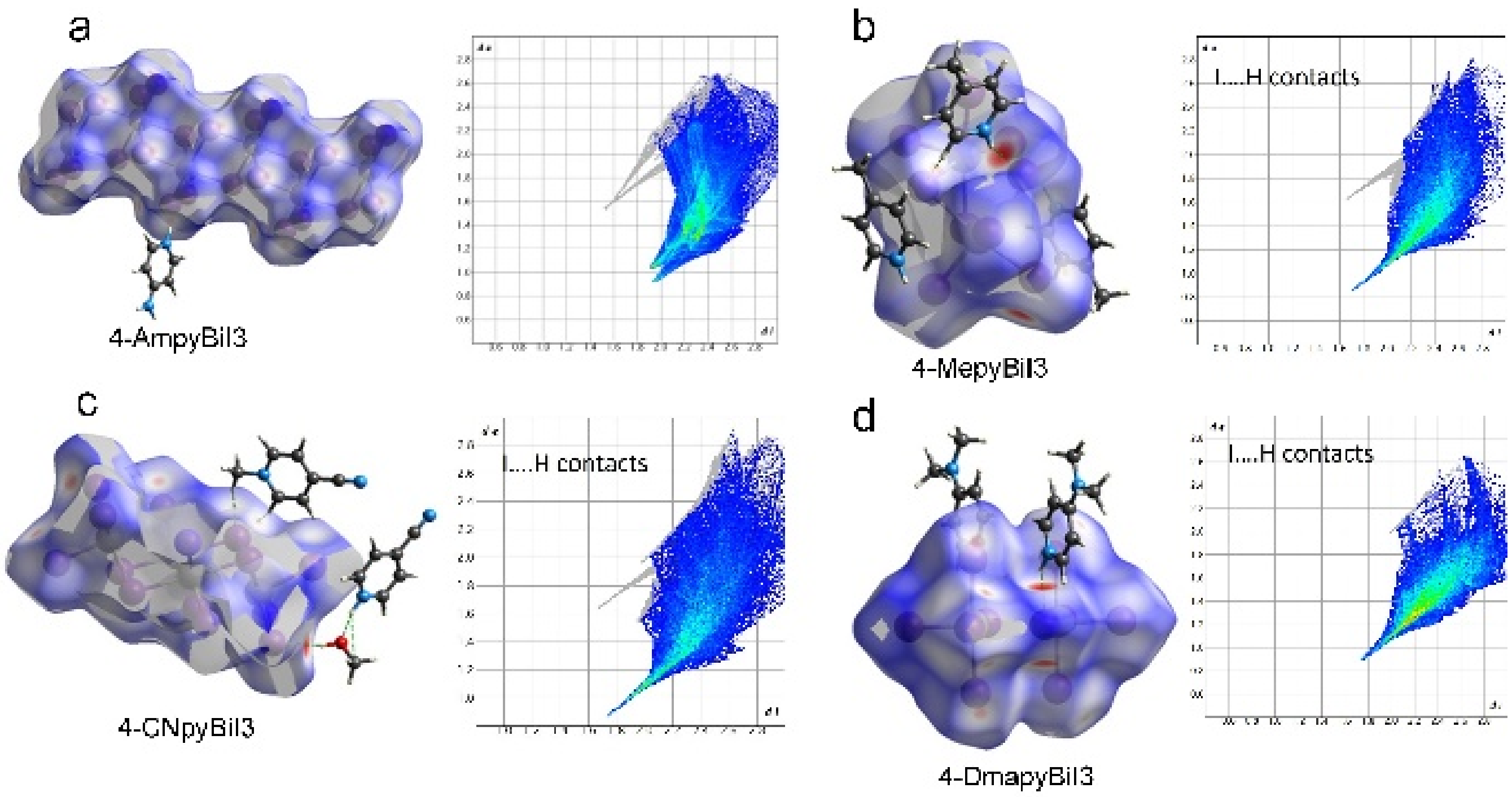

***Figure 4**. 3D HS plotted as $d_{norm}$ in and 2D finger print of I···H interactions of 4-AmpyBiI$_3$ (a), 4-MepyBiI$_3$ (b), 4-CNpyBiI$_3$ (c) and 4-DmapyBiI$_3$ (d).*

***Figure 5***. *Percentage of intermolecular interactions in 2D* $d_{norm}$ *analysis.*

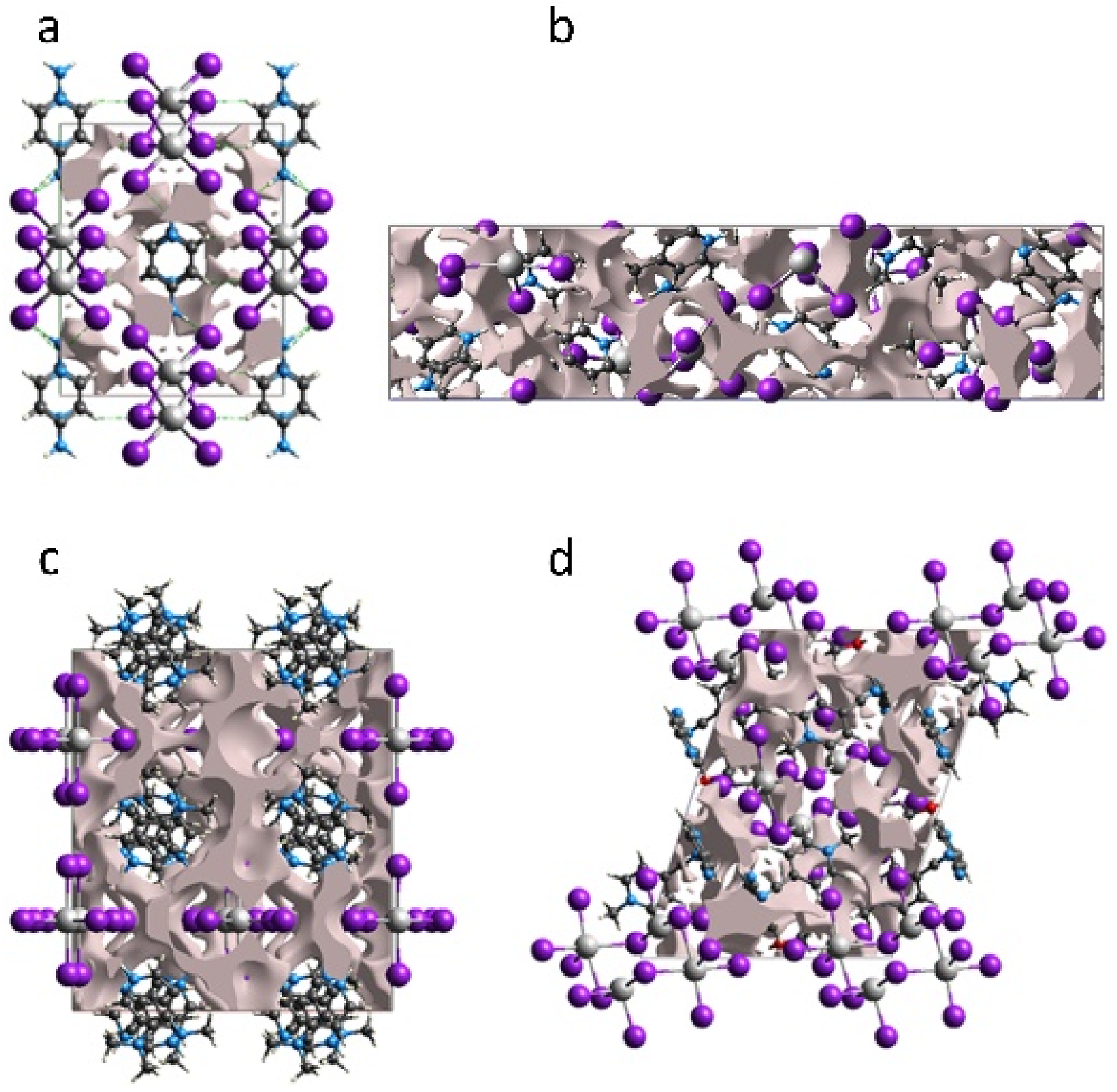

***Figure 6***. *Depicts the void spaces in the crystals unit cells of the (a) 4-Ampy*$BiI_3$*, (b) 4-Metpy*$BiI_3$ *(c) 4-Dmapy*$BiI_3$*, (d) 4-CNpy*$BiI_3$ *crystals.*

In addition to the above, the two-dimensional (2D) fingerprint plots offer quantitative data on total contact and each interatomic interaction in the area of the HS (see Figure 4, Figure SI6). When calculating individual interatomic connections, reciprocal interactions are also factored in. Further investigation through the 3D map of $d_{norm}$ and the 2D plot for the I···H interactions, reveals their significant influence on crystal packing, as they contribute more than 80 % in case of 4-AmpyBi$I_3$, 4-MepyBi$I_3$,4-DmapyBi$I_3$) (Figure 5). Following these, I···I, and I···C interactions also account for a considerable percentage, especially last one in case of 4-CNpyBi$I_3$ has a high value (16.9%).

CrystalExplorer was used to visualize voids within each unit cell of the crystalline structures using electronic density isosurfaces (Figure 6).[40] The void percentages are defined by the sum of the spherical electron densities at the corresponding sites. A unit cell with a low void volume or percentage of vacancy typically indicates that the molecules are tightly packed together.

The surface area, globularity, and asphericity of each structure is provided in Table 4. Among these, the 4-AmpyBi$I_3$ structure with 1D fragments ($[BiI_4]_n^{n-}$) has the lowest void volume, constituting only 4.3% of its unit cell volume. Both globularity and asphericity are metrics derived from the Hirshfeld analysis of volume and surface area. Molecular anisotropy is measured by asphericity, whereas globularity is the difference between the surface area calculated by the Hirshfeld method and the sphere area for a given volume.[41] Further details on the Hirshfeld calculations and void analysis are provided in Table 4 and the Supplementary Information (Figure SI6). Interestingly, 4-MepyBi$I_3$ which has a non-centrosymmetric crystalline space group, displays a higher asphericity value (0.658 Ω), while the others approximate a value near 0, indicative of the isotropic characteristic.[42]

### 3.4 Density functional theory (DFT) calculations

As the electrical conductivity and mobility of photogenerated electrons (or holes) play a crucial role in the optoelectronic device's performance, in this part DFT calculation of the synthesised structures was considered. The effective masses of CB electrons ($m_e$*) and VB holes ($m_h$*) were calculated using the band structures shown in Figure 7, Figure SI 7, from the second derivative of the energy state ($E(k)$) versus the crystal momentum $k$ using the following equations (2-3):

$$\frac{1}{m_h^*} = \frac{1}{\hbar^2}\frac{d^2 E_{VB,max}}{dk^2} \qquad (2)$$

$$\frac{1}{m_e^*} = \frac{1}{\hbar^2}\frac{d^2 E_{CB,min}}{dk^2} \qquad (3)$$

Table 5 provides the effective masses at the $VB_{max}$ and $CB_{min}$ edges. For 4-AmpyBi$I_3$, the effective masses of both the electron and the hole are relatively low, implying potentially high charge mobility. For the remaining three structures, the effective mass of the hole ($m_h$*) is smaller than that of the electron ($m_e$*). This difference could be attributed to the characteristics of the semiconductors, since the curvature at the bottom of $VB_{max}$ is less pronounced than at $CB_{min}$.

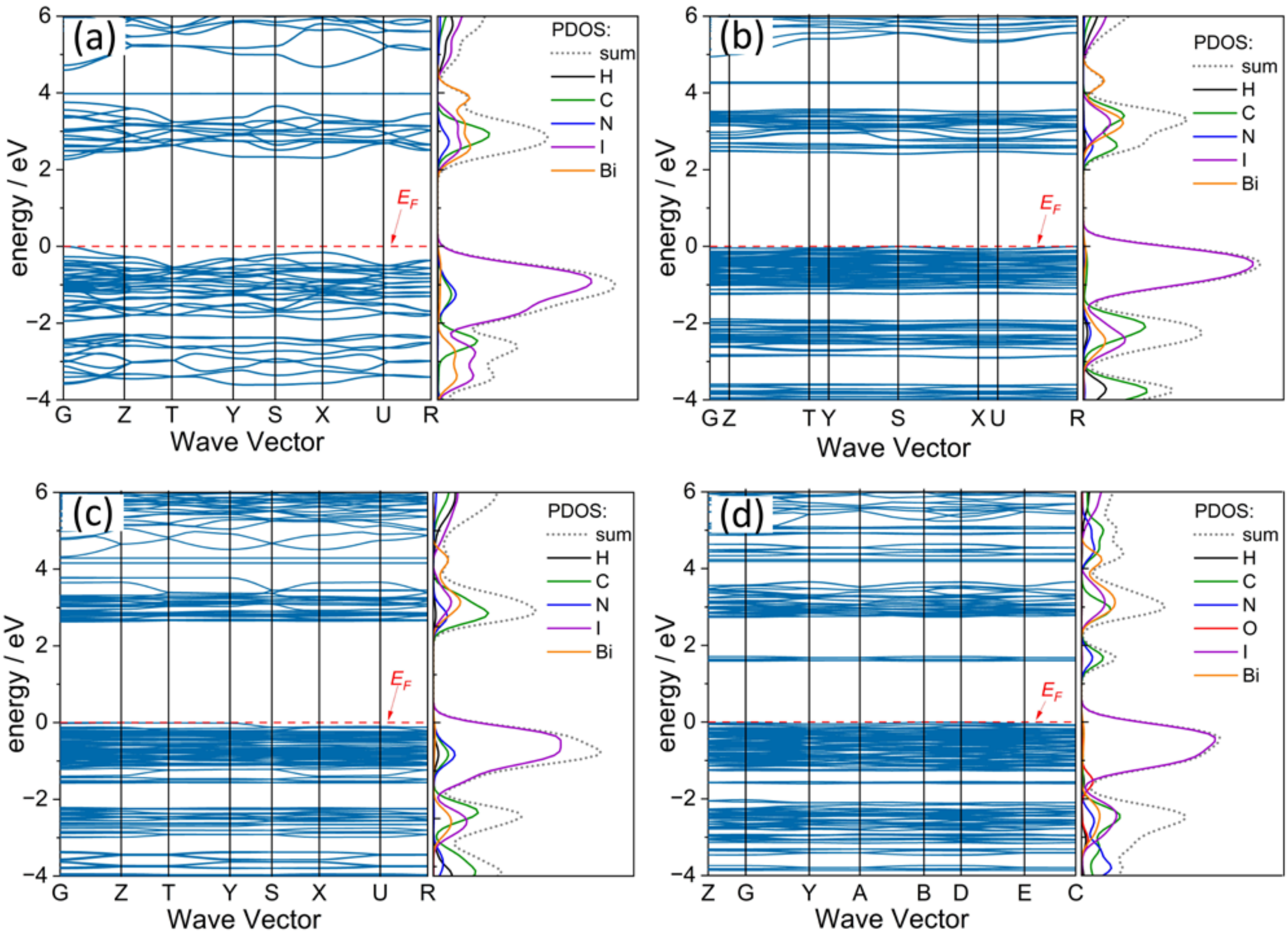

***Figure 7.*** *Left side: DFT-calculated band structures of (a) 4-AmpyBiI$_3$, (b) 4-MepyBiI$_3$, (c) 4-DmapyBiI$_3$, and (d) 4-CNpyBiI$_3$ collated with the right side: partial density of states (PDOS).*

The band structure diagrams along with the DOS and PDOS distributions are shown in Figure 7. Except for 4-AmpyBiI$_3$, the dispersion diagrams are very flat, which is characteristic of ionic compounds with high effective masses of charge carriers. This reflects weak intermolecular electronic coupling and is a consequence of disruption[43] of the 2D structure of the parent bismuth iodide[44, 45] into isolated bismuth iodide clusters.[43] The 4-AmpyBiI$_3$ derivative stands out as it retains a partially preserved covalent network (1D bismuth iodide chains), therefore exhibiting the larger dispersion and lower corresponding effective mass of charge carriers. Interestingly, there appears to be no direct correlation between the dispersion relation and projected DOS distributions. All compounds, except 4-CNpyBiI$_3$, show similar PDOS profiles. The valence band is mostly composed of 6s and 6p orbitals of iodide anions, with minor contributions from carbon 3p and bismuth 6s orbitals, especially in deeper regions of the valence band. In the case of 4-MepyBiI$_3$ and 4-CNpyBiI$_3$ the valence band is solely based on iodine orbitals. Valence bands are composed of 6p orbitals of bismuth, 3p of carbon, 3p of nitrogen, and 6p of iodide. The 4-CNpyBiI$_3$ again differs, with the lower part of the conduction band containing only 3p orbitals of carbon and nitrogen, and the second conduction band (approx. 1 eV above the first one) is hybridised from orbitals of bismuth, carbon, and iodine.

**Table 5.** *Parameters calculated from band structures of bismuth-based complexes.*

| Entry | Compound | $E_g$, exp [a] | $E_g$, th [b] | $m^*_h$ [c] | $m^*_e$ [d] |
|---|---|---|---|---|---|
| 1 | 4-AmpyBi$I_3$ | 1.91 | 2.27 | 1.1 | 0.7 |
| 2 | 4-MepyBi$I_3$ | 1.86 | 2.42 | 4.4 | 6.1 |
| 3 | 4-DmapyBi$I_3$ | 1.99 | 2.63 | 25 | 49 |
| 4 | 4-CNpyBi$I_3$ | 1.73 | 1.59 | 7.3 | 32.4 |

*a: experiment (DRS method)*

*b: theory (DFT method)*

*c: effective hole mass, relative to the free electron mass, $m_0$*

*d: effective electron mass, relative to $m_0$*

This deviation can be linked to the strong electron-withdrawing character of the CN moiety, which results in the 4 cyanopyridinium and 4 cyano-N-methylpyridinium cations being strong electron acceptors. As a result, the conduction band is solely comprised of cationic sublattice, a unique feature in this family of compounds. This is also associated with the lowest band gap energy (1.73 eV) observed for this compound. Other band gap values and additional electronic parameters of these materials are collated in Table 5.

*Optical Properties*

The pyridinium iodobismuthates under investigation are insoluble in water, but they dissolve well in several common polar organic solvents. UV-vis absorption spectra of the prepared samples were measured in two aprotic solvents: a coordinating solvent (dimethylformamide, DMF) and a non-coordinating one (propylene carbonate, PC), as shown in Figure 8. Along with the spectra of studied compounds, the spectra of bismuth iodide were recorded. Bismuth iodide shown absorption peaks at 425 nm in DMF and 481 nm in PC. The spectra of iodobismuthates are almost identical both in both DMF (ranging from 425-432 nm, depending on the cation) and PC (all compounds showed a peak at 475 nm). This indicates that the compounds undergo almost complete dissociation upon solvation and the observed spectral features are only observed in monomeric iodobismutate moieties, $[BiI_4]^-$ and $[BiI_6]^{3-}$.[46] Characteristic spectral features associated with polymerised iodobismuthate anions are not observed.[47] The higher energies of the lowest transition with ligand-to-metal charge transfer (LMCT) character could be due to partial coordination of DMF to bismuth centres and/or solvatochromic effect due to distortion of the octahedral symmetry of the complex.

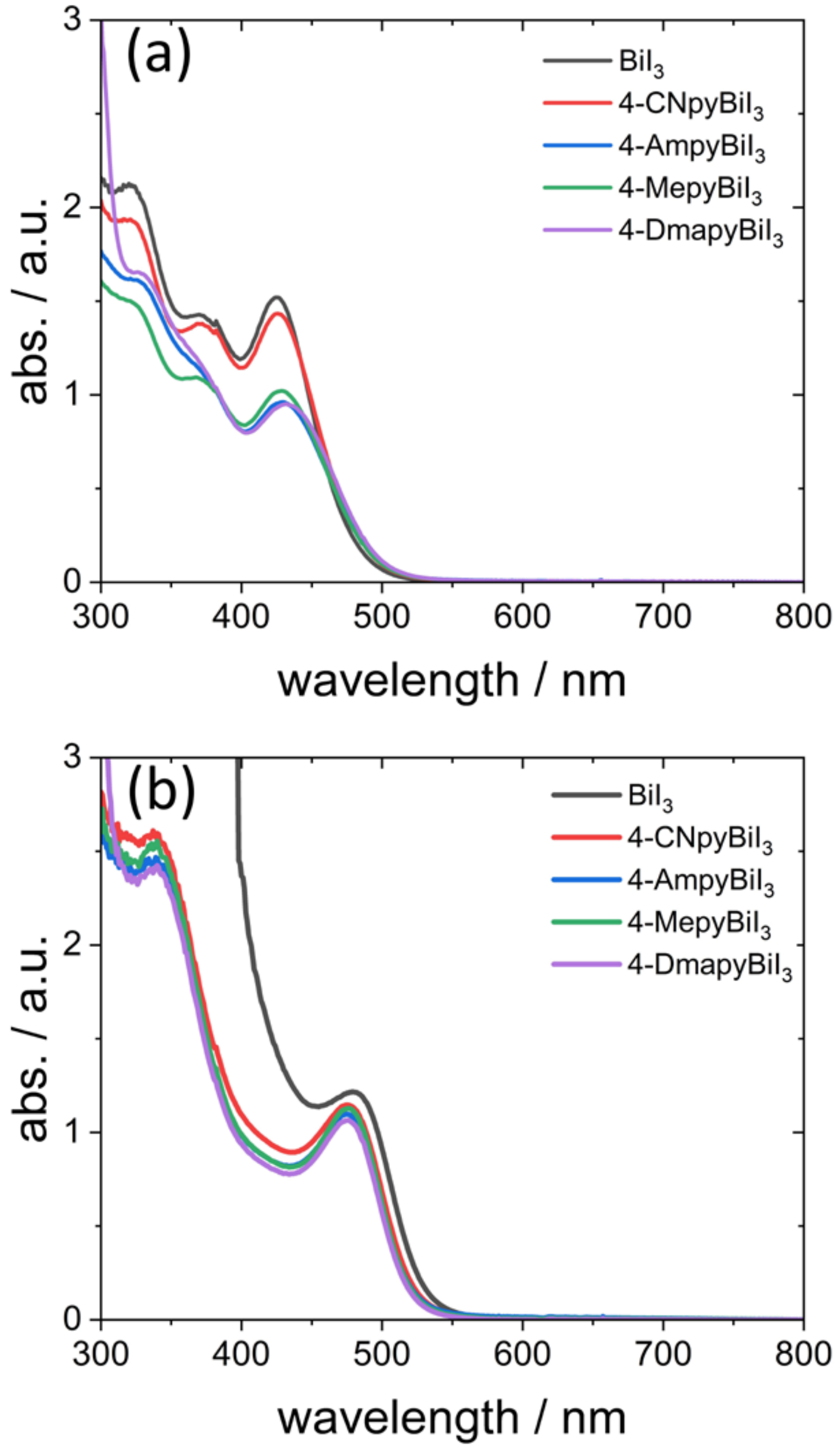

***Figure 8.*** *UV-vis absorption spectra of various pyridinium iodobismuthates compared with the bismuth iodide in coordinating (dimethylformamide, a) and non-coordinating (propylene carbonate, b) solvents.*

In contrast, the diffuse reflectance spectra show a striking difference to the solution spectra (Figure 9). No distinct molecular features associated with individual iodobismuthate moieties are observed.[47] Instead, the absorption onset characteristic of a large band gap semiconductor is evident.[48] This observation highlights the dominant role of intermolecular interactions (vide supra) on the electronic structure and optical properties of the studied compounds. Similar effects have been observed for other ionic and molecular semiconductors.[20, 21] The band gap values can be derived from a linear fit of the Tauc equation (4):

$$(\alpha h\nu)^n = A^*(h\nu - E_g) \quad (4)$$

In this equation, $\alpha$ represents the absorption coefficient (here, the Kubelka-Munk function), $A^*$ is a parameter dependent on the refraction coefficient and the effective masses of the charge carriers, $E_g$ is the band gap energy, and $n$ is a parameter describing the type of fundamental transition.[48] The derived band gap values are summarised in Table 5. It is evident that careful selection of the cation can significantly modulate the band gap of resulting semiconductors. Specifically, strong electron-withdrawing groups cause complete rearrangement of the band structure, as already discussed for the case of 4-CNpyBi$I_3$ case (*vide supra*). Moreover, hybridisation of 6s and 6p orbitals results in high polarisability of the [Bi$I_6$] core, slight differences in local electron density distribution and dipole moments can significantly affect the polymerisation and distortion of iodobismuthate octahedra. Given the polarisability, even weak hydrogen bond interactions are sufficient to promote significant mixing of 6s and 6p orbitals of bismuth with the 3p orbitals of carbon and nitrogen in the conduction band. This affects not only the composition of bands but also the LUMO/HOMO energy, which is reflected in the modulation of the band gap energy (Table 5).[49] The determination of the band gap energies and transition character with precision is not straightforward. The DOS plots indicate a direct bandgap for 4-AmpyBi$I_3$. However, because of the very flat dispersion for other materials, the band gap cannot be unequivocally defined. Therefore, the three possible form of the Tauc equation were applied: for direct ($n$ = 2, Figure 9a) and indirect ($n$ = ½, Figure 9b) transitions, alongside with the equation characteristic for molecular and amorphous semiconductors ($n$ = 1, Figure 9c). In the first two cases good quality linear dependencies were found, and the corresponding band gap energies were calculated. Furthermore, due to the negligible overlap of electron density between individual ions in the lattice, a third equation for molecular and amorphous semiconductors was used ($n$ = 1, Figure 9c). Due to the structure and nature of bonding within the lattices of the compounds studied, this approach seems to be the most justified, as has been the case with other similar materials.[20, 21]

*X-ray absorption spectroscopy (XAS).*

X-ray absorption spectroscopy (XAS) has been used to examine the electronic structures of the complexes (Figure 10), specifically utilising the Bi $L_3$-edge. Because of the core-hole shielding effect, the oxidation state of an element can be ascertained from the shift in the location of the absorption edge. Electron-nucleus attraction and electron-electron repulsion define the potential energy of an electron. As the number increases, the atom loses more electrons from its shells, which weakens the nucleus shielding. Consequently, a higher amount of energy is needed to remove the subsequent electron, a change that is reflected in the shift in the absorption edge towards higher energy levels.

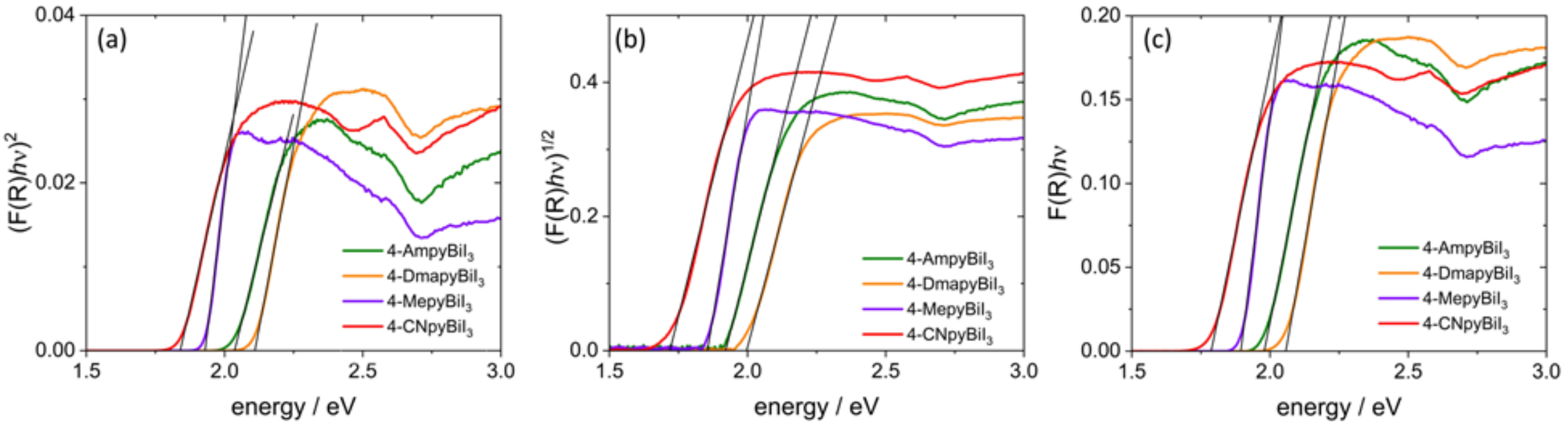

***Figure 9**. UV-vis diffuse reflectance spectra (Tauc plots derived from Kubelka-Munk function) for direct (a) and indirect (b) transitions, as well as calculated for molecular semiconductors (c).*

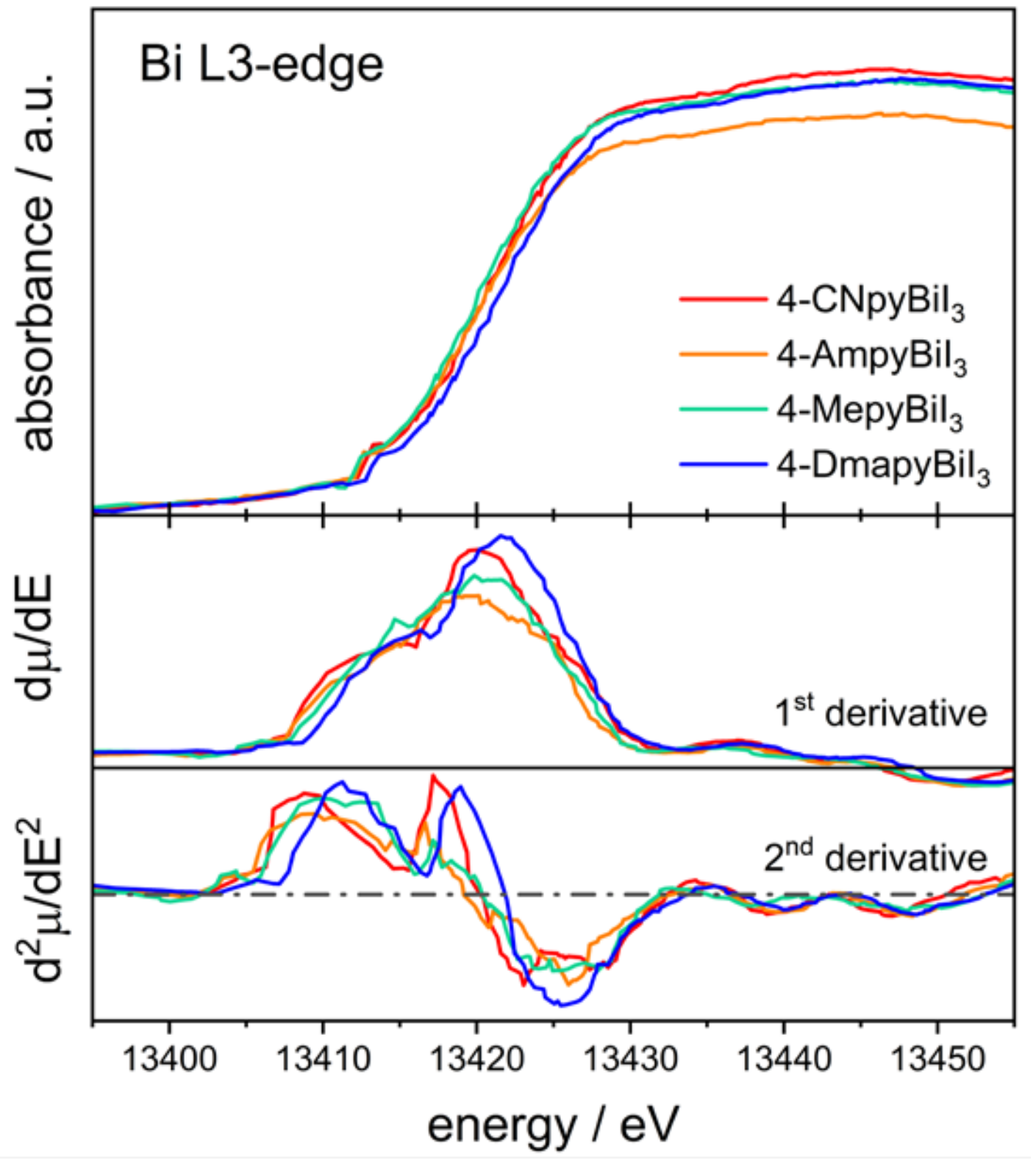

***Figure 10.** Experimental XAS spectra for 4-AmpyBiI$_3$, 4-MepyBiI$_3$, 4-DmapyBiI$_3$, and 4-CNpyBiI$_3$, complexes at the L3 edge are presented on top, together with their first (middle) and second (bottom) derivatives. The dotted grey line is on the 2$^{nd}$ derivative plot and stands for 0. The derivatives were additionally smoothed using a Savitzky-Golay filter.*

Figure 10 shows the X-ray Absorption Near Edge Structure (XANES) at the $L_3$-edge measured for the complexes which characteristically aligns with $Bi^{3+}$. The spectrum consists of a simple peak with a 'delayed' maximum exceeding the absorption threshold.[50] The position of the absorption edges, often determined as the maximum of the first derivative of the spectra, can also be estimated as the energy at half of the height of the absorption step. The absorption edges are very close for all the studied materials, around 13421 eV. This indicates that the oxidation state and the ligand environment of bismuth in the studied compounds remain the same. The recorded spectra are noticeably different from those reported for $BiI_3$ and $Cs_3Bi_2I_9$ by Phuyal et al.[51] These subtle changes seem to reflect the differences in hybridisation between the 6s and 6p orbitals of bismuth and ions within the [$BiI_6$] octahedra, which are involved in different interaction manifolds. Interestingly, the absorption edges of the studied bismuth complexes (approximately 13421 eV) are significantly higher than those reported for $BiI_3$ (approximately(approximately 13416 eV) or $Cs_3Bi_2I_9$ (ca 13416.5 eV). The polarisability of the [$BiI_6$] motif should disturb only the local symmetry, without affecting the energy of the absorption edge. Therefore, the discrepancy in the absorption edge values may be due to the energy calibration method.

The position of the absorption edge only gives a general insight into the oxidation state or the ligand field around the absorber. More information can be found by analysing the absorption features both below and above the main edge. This was carried out using FDMNES code, a tool that determines how particular electronic transitions affect the fine structure of the XAS spectra. Figure SI8 shows that the convoluted DFT-calculated spectra closely resemble the general shape of the experimental data. In addition, the partial density of the unoccupied electronic states (DOS) corresponding to the calculated curves is provided. Importantly, FDMNES calculates the DOS projected on the absorbing atom in relation to the absorption process (energy) by solving the Schrödinger equation in the one-electron approximation. Thus, DOS is influenced by the different molecular orbitals that contribute to the spectral profiles at the probed edge $L_3$, rather than at the actual probed edge itself. The plotted DOS is a sum of all the principal quantum numbers of the particular orbital projected on the angular momentum.

The basic electron configuration of the bismuth atom is [Xe] $4f^{14}5d^{10}6s^26p^3$. Depending on the oxidation state, it may have valence electrons on 6s and 6p orbitals. The absorption around the bismuth $L_3$-edge mainly from the excitation of $2p_{3/2}$ electrons to the dipole-allowed s- and d-type orbitals. However, a less likely quadrupole-allowed transition to p- and f-type orbitals also exists. The corresponding selection rules stands as $\Delta l = \pm 1$ for dipolar transitions and $\Delta l = 0, \pm 2$ for quadrupolar transitions, where l is the quantum number. No ‘white line’ is observed because its intensity for the $L_3$ edge is often proportional to the number of unoccupied d orbitals, which are fully occupied for the Bi. Bismuth in the complexes studied should reveal the +3 oxidation state with valence electrons only in the 6s orbital.[51] This is confirmed by the lack of a

*Electric Properties*

4-subsdtituted pyridinium iodobismutates share structural and electronic features with halide perovskites. Although these materials have distinct lattice structures, they are based on partially fused [$BiI_6$] octahedra, leading to 1D (4-aminopyridinium) or 0D (other cations) motifs. The electronic structure (Figure 7 band structure), as well as optical spectroscopy, indicate a clear semiconducting character of these materials, implying that they should show measurable electronic conductivity. Furthermore, the presence of large voids in the lattice and a complex network of hydrogen bonds, combined with high polarisability of the [$BiI_6$] unit, suggests the potential foe ionic conductivity along with electronic one.

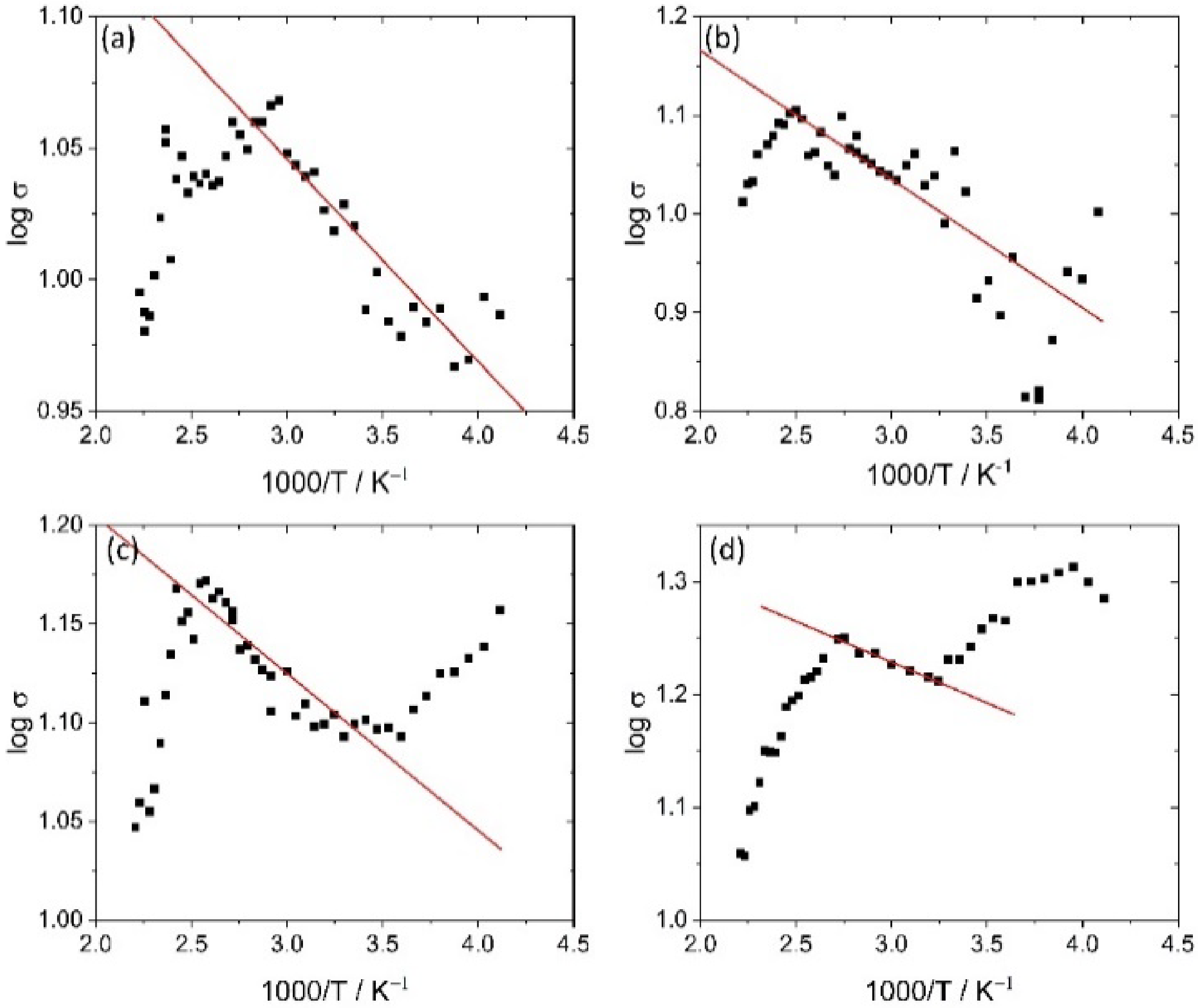

***Figure 11.*** *Arrhenius plots for electrical conductivity of 4-AmpyBiI₃ (a), 4-DmapyBiI₃ (b), 4-MepyBiI₃ (c), and 4-CNpyBiI₃ (d). The red line is a linear fit to the Arrhenius equation.*

The linear sweep voltammetry (LSV) technique was applied to measure the current-voltage characteristic of devices made of thin layers of the prepared materials on ITO/glass. This revealed conductivity values within the range of 10-20 mS/pixel at 25°C, the highest recorded for 4-CNpyBi$I_3$, and the lowest for 4-AmpyBi$I_3$. These observations do not correlate with the calculated effective masses of charge carriers (as presented in Table 5), suggesting that factors other than the electronic structure of the materials must contribute to the transport phenomena in pyridinium iodobismuthates. An obvious factor for this influence would be ionic conductivity, especially the migration of iodide anions and lattice defects in applied electric fields. In order to gain a better understanding of the conductivity mechanism, the temperature dependence of conductivity was evaluated. The resulting Arrhenius plots (Figure 11) show two distinct regions for the two compounds 4-AmpyBi$I_3$, and 4-DmapyBi$I_3$: a low temperature region, where conductivity increases with temperature, and a high-temperature region, in which conductivity decreases with increasing temperature. This behaviour is consistent with combined electric and ionic conductivity. Arrhenius plots indicate activation energies for ionic conductivity at 76 and 130 meV for 4-AmpyBi$I_3$, and 4-DmapyBi$I_3$, respectively. Although these values are comparable, they are significantly lower than the corresponding activation energies for lead iodide perovskites (200-400 meV).[52] Lower activation energies are associated with greater flexibility of the lattice, the presence of larger organic cations, and complex void structures, allowing for facile ionic movements. At a higher temperature, however, lattice oscillations efficiently scatter migrating ions, thus resulting in decreased conductivity.

The methyl- and cyanopyridinium derivatives show even more complex thermal dependence of the conductivity. In the low-temperature regime, conductivity decreases with increasing temperature, followed by an increase and a subsequent decrease of conductivity, similar to the previous observations. The activation energies for ionic transport are comparable to those of previous cases, with 72 and 80 meV for the 4-cyanopyridinium and 4-methypyridinium complexes, respectively. These findings are consistent with previously reported data for lead halide perovskites and other pyridinium iodobismuthates.[52] Furthermore, the I-V curves (Figure SI9) reveal a significantly non-linear character, further supporting the hypothesis of mixed electronic and ionic transport at the studied junctions. Moreover, these junctions show moderate rectifying character with current ratios for forward-to-reverse polarisation amouting ca. 1.5 to 2.2.

**Conclusions**

In this study, pyrridinium iodobismuthates are comprehensively examined, highlighting their potential as a promising class of semiconductors with applications in photovoltaics and memristive devices. They exhibit thermally stable crystal lattices at least up to 230°C, when pyridine is functionalised by electron donating groups (-methyl, -amino, and -dimethyamnino). Moreover, the family shows tunable bandgaps and complex interplays of electronic and ionic conductivity, depending on the type of derivatives and crystallographic dimensionality. Furthermore, they present an intriguing platform for exploring crystal engineering, showing how subtle changes in cation geometry may affect the crystal structure and electronic properties of the ionic compound. The good solubility in organic solvents and stability in a humid environment enhance their potential applications, presenting them as strong contenders to replace the less stable lead halide perovskites in certain applications. While their transport properties and relatively low absorption coefficient may render them unsuitable for photovoltaic applications, they display promising characteristics for memristive and optoelectronic properties, drawing parallels with related structures of aromatic amines directly bound to bismuth centres. This hypothesis can be further justified by their weak rectifying behaviour (possibly due to the formation of a leaky Schottky barrier at the metal-complex interface), the presence of telegraphic noise-like features in current-voltage characteristics,[53, 54] and their structural relationships with other materials, with well-established resistive switching characteristics.[19, 55, 56] These findings shed light on the untapped potential of these compounds and open avenues for further research and development in the field of semiconductor technology.

## Acknowledgement

The authors acknowledge the financial support of the Polish National Science Centre within the OPUS programme (grant agreement No. UMO-2020/37/B/ST5/00663). This research was partially supported by the programme 'Excellence initiative-research university' for the AGH University of Science and Technology and by the NAWA programme (contract No. PPN/BIN/2019/1/00100). The DFT calculations have been performed at the CYFRONET AGH Academic Computing Centre with computational grant PLGSURFCACE4.

## Authors' Contributions

GA: Methodology, investigation, data curation, formal analysis, visualisation, writing - original draft, review and editing; MG: methodology, investigation, data curation, formal analysis, writing – original draft, review and editing; AS: methodology, investigation, data curation, formal analysis, visualization, writing – original draft; EK: investigation; TM: investigation, writing – original draft; AP: investigation; KM: investigation; PZ: methodology; AP: investigation; AK: investigation; AA: investigation; CV: writing – original draft, review and editing; KS: conceptualisation, methodology, visualization, funding acquisition, formal analysis, writing – original draft, review and editing.

## Conflicts of interest

There are no conflicts to declare.

## Notes and references

# Supporting Information

***Table SI 1***. *Elemental analysis of obtained pyridinium iodibismuthates.*

| Compound | Mass average[1]/C:H:N | Mole ratios (Experimental result)[1] | Mole ratios (Theoretical result)[2] |
|---|---|---|---|
| 4-AmpyBiI$_3$ | 9.14:3.73:0.98 | 5.71:7.24:2 | 5:7:2 |
| 4-MetpyBiI$_3$ | 8.84: 0.96: 1.62 | 6.37:8.24: 1 | 6:8:1 |
| 4-DimetpyBiI$_3$ | 14.49: 1.9: 4.68 | 7.2:11.25: 2 | 7:11:2 |
| 4-CNpyBiI$_3$ | 9.46:3.21:0.861 | 6.88:7.4:6:2 | 6:5:2 |

[1]Elemental analysis

[2]Protonated organic moiety

***Table SI2.*** *Summary of TG-DTG results.*

| Compound | Δm [%] | $T_{decomposition}$ | | |
|---|---|---|---|---|
| | | $T_{start}$ , [°C] | $T_{max,}$ [°C] | $T_{end,}$ [°C] |
| 4-AmpyBiI$_3$ | 99.98 %<br>(24 °C – 500 °C) | 236.4 | 400.2 | 405.7 |
| 4-MepyBiI$_3$ | 99.81%<br>(24 °C – 500 °C) | 231.1 | 371.8 | 378.3 |
| 4-DmapyBiI$_3$ | 93.93%<br>(24 °C – 500 °C) | 282.9 | 416.5 | 449.5 |
| 4-CNpyBiI$_3$ | 23 %<br>(24 °C – 246°C) | 93.3 | 239.7 | 245.7 |
| | 76.98 %<br>(246 °C - 400 °C) | 246 | 350 | 400 |
| pyBiI$_3$ | 26.88%<br>(24 °C – 159.03°C) | 78 | 141 | 159.03 |
| | 73.34%<br>(159.03 °C – 274.98°C) | 159.03 | 233.89 | 274.98 |

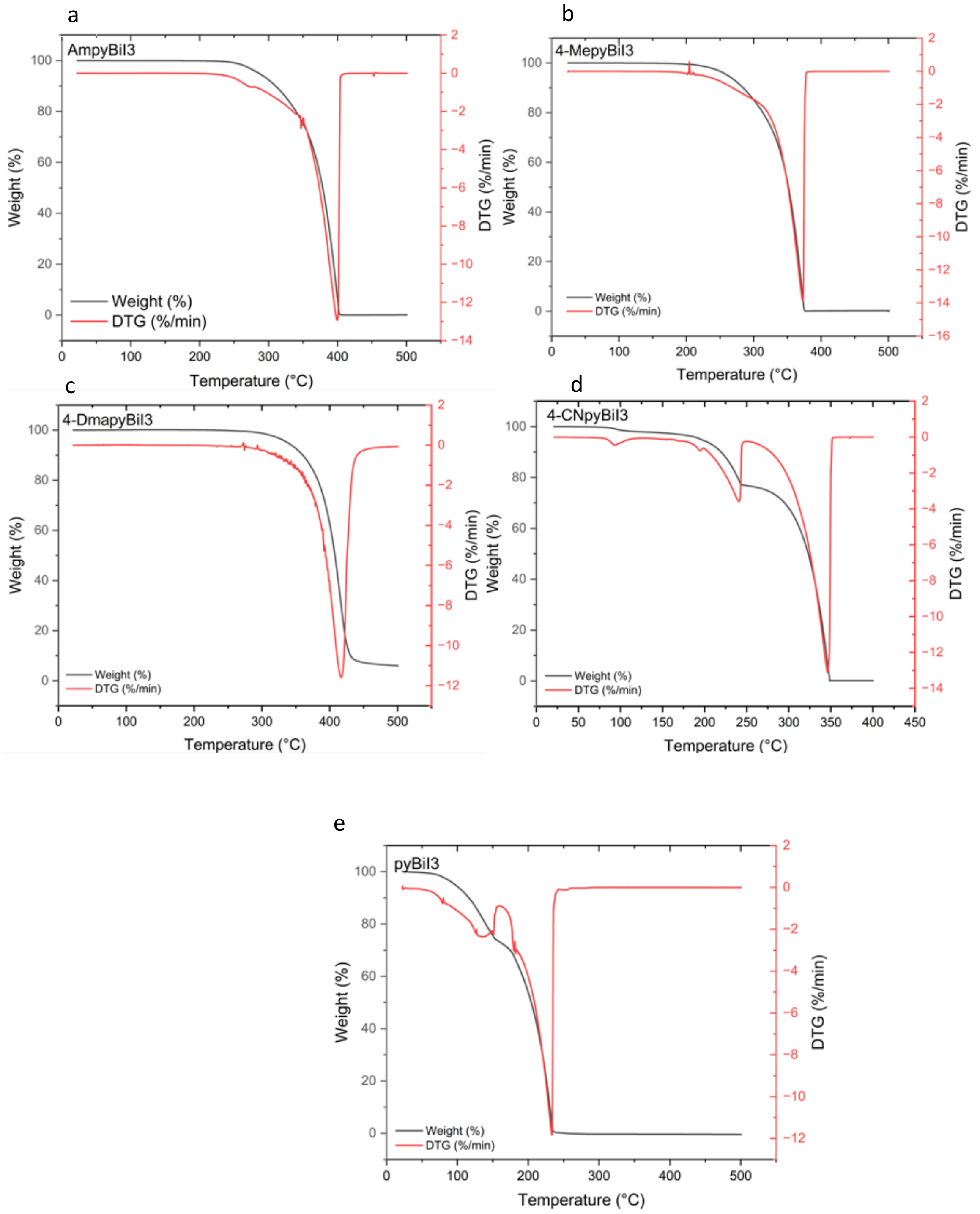

***Figure SI 1***. *TG-DTA of a) 4-AmpyBi$I_3$ b) 4-MetpyBi$I_3$ c) 4-DmapyBi$I_3$ d)4-CNpyBiI3 e) pyBi$I_3$.*

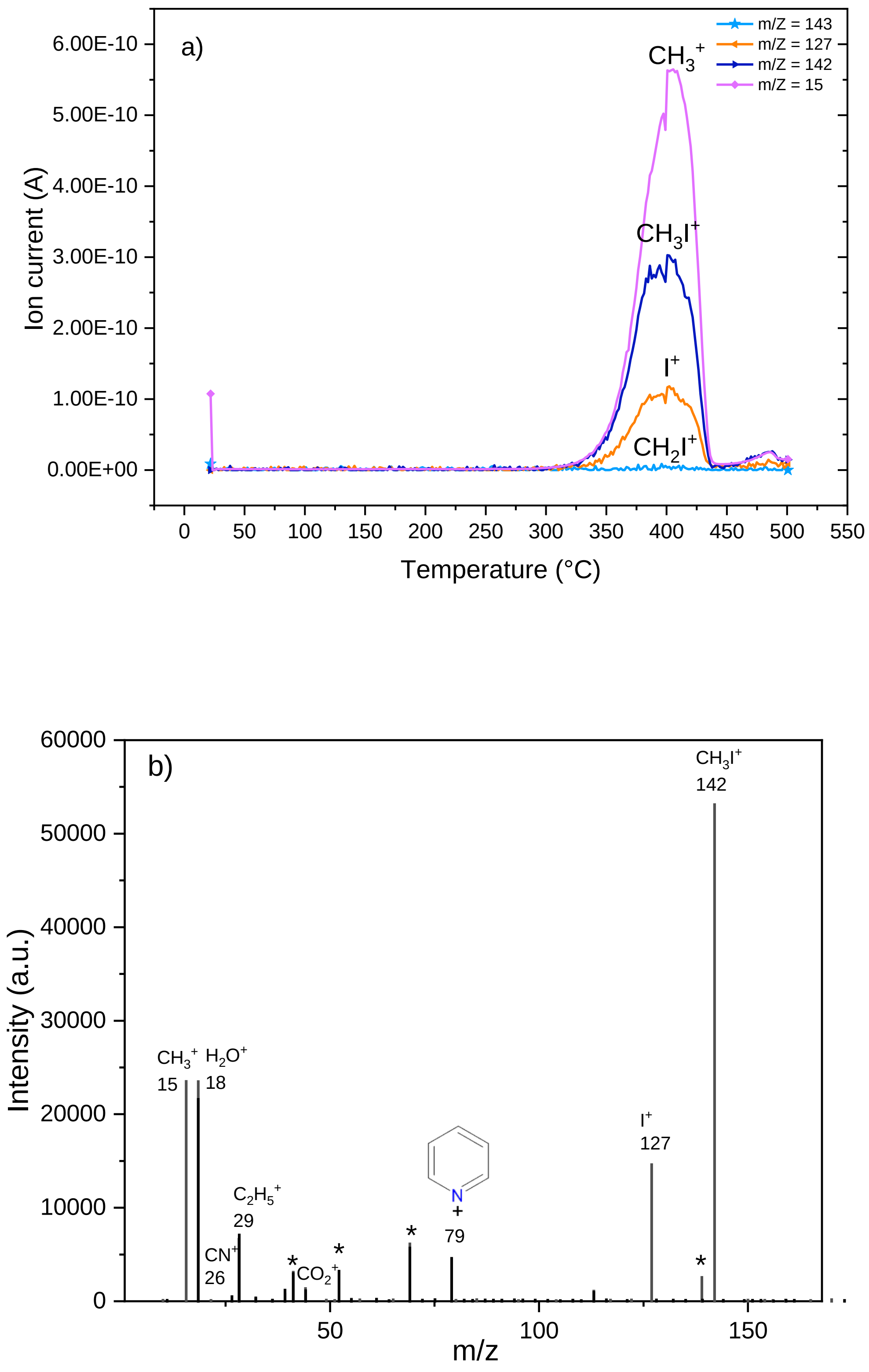

***Figure SI 2**. (a) TGA-Mass spectrum of 4-DmapyBiI$_3$, (b) GC-Mass spectrum of 4-MepyBiI$_3$.*

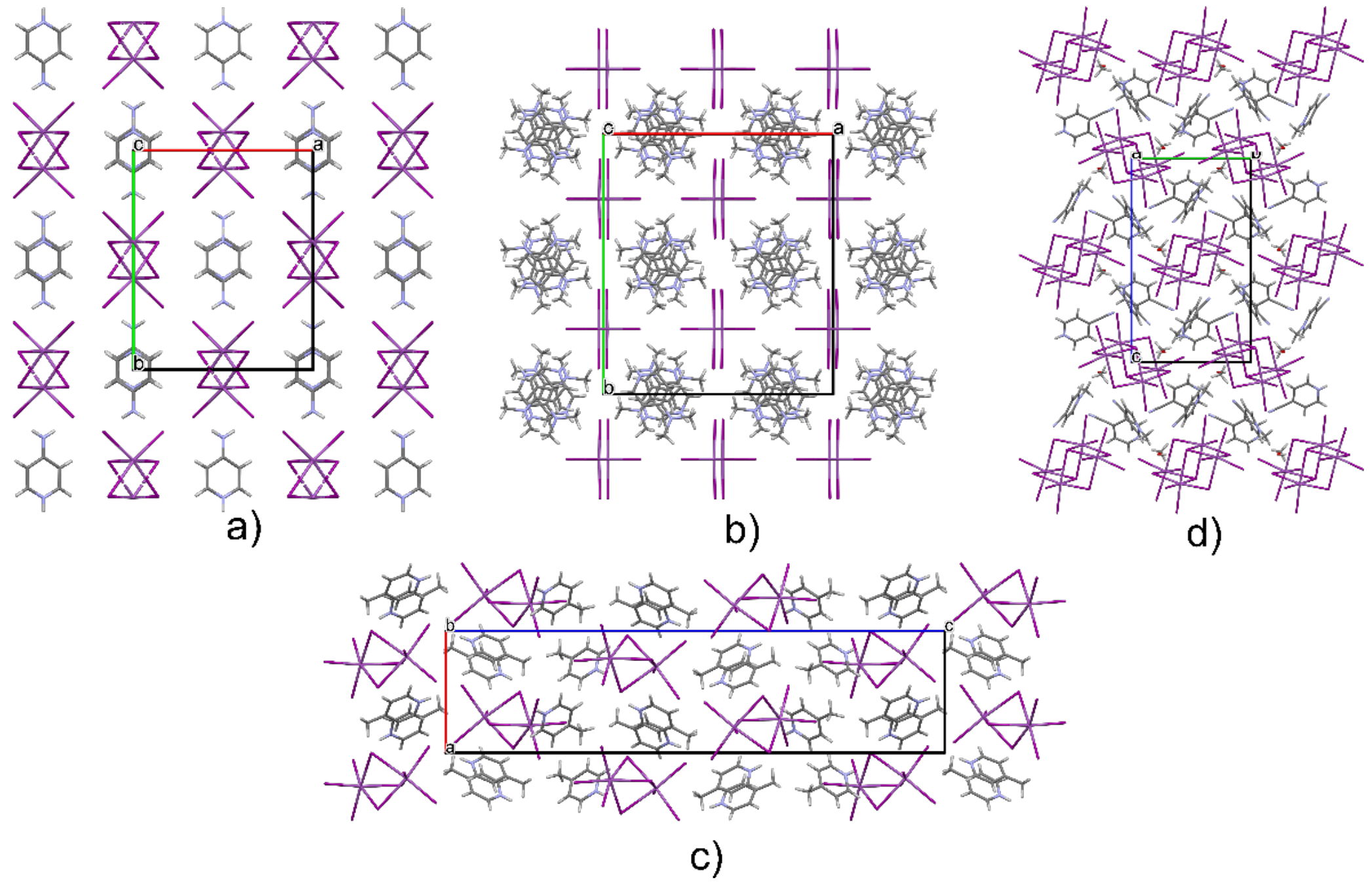

**Figure SI 3.** Packing of the structural components in a) 4-Ampy$BiI_3$ along [001], b) 4-Dmapy$BiI_3$ along [001], c) 4-CNpy$BiI_3$ along [010], d) 4-Mepy$BiI_3$ along [100].

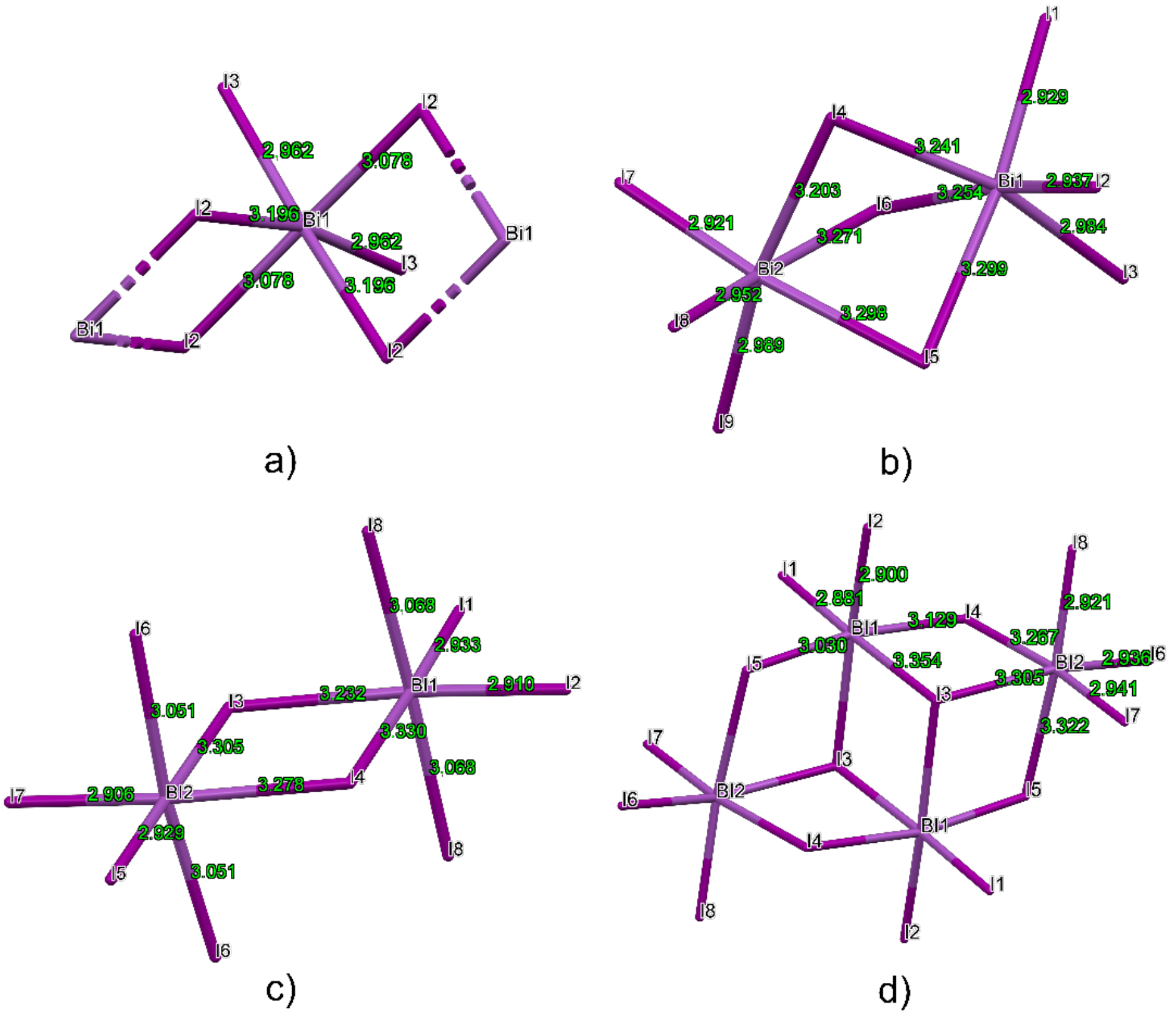

**Figure SI 4.** a) $[BiI_4]_n$ in 4-Ampy$BiI_3$ along [001], b) $Bi_2I_9$ in 4-Dmapy$BiI_3$ along [001], c) $Bi_2I_{10}$ in 4-CNpy$BiI_3$ along [010], d) $Bi_4I_{16}$ in 4-Mepy$BiI_3$ along [100].

***Table SI3***. *Geometry of weak interactions in 4-AmpyBiI$_3$ (1) [Å/°]*

| D-H···A | d(D-H) | d(H···A) | d(D···A) | <(DHA) |
|---|---|---|---|---|
| N(8)-H(8)...I(3) | 0.88 | 3.06 | 3.715(3) | 133.0 |
| N(8)-H(8)...I(3)#1 | 0.88 | 3.06 | 3.715(3) | 133.0 |
| N(4)-H(4A)...I(3)#4 | 0.87(2) | 2.93(5) | 3.7242(18) | 153(9) |
| C(7)-H(7)...I(2)#5 | 0.95 | 3.31 | 3.805(3) | 114.5 |
| C(5)-H(5)...I(2)#6 | 0.95 | 3.10 | 3.947(3) | 149.7 |

Symmetry transformations used to generate equivalent atoms:
#1 -x+1,y,-z+3/2 #2 -x+1,-y+2,-z+1 #3 x,-y+2,z+1/2
#4 -x+1,-y+1,-z+2 #5 -x+1/2,-y+3/2,z+1/2 #6 -x+1/2,y-1/2,z

***Table SI4***. *Geometry of weak interactions in 4-DmapyBiI$_3$ [Å/°]*

| D-H···A | d(D-H) | d(H···A) | d(D···A) | <(DHA) |
|---|---|---|---|---|
| C(3A)-H(3A)···I(8)#2 | 0.95 | 3.08 | 3.965(3) | 155.8 |
| C(6A)-H(6A)···I(6)#3 | 0.95 | 3.06 | 3.723(3) | 128.6 |
| C(6A)-H(6A)···I(8) | 0.95 | 3.29 | 3.827(3) | 118.0 |
| C(8A)-H(8A2)···I(1)#4 | 0.98 | 3.26 | 4.110(4) | 146.6 |
| N(1A)-H(1A)···I(3) | 0.88 | 2.92 | 3.702(3) | 148.2 |
| N(1B)-H(1B)···I(4) | 0.88 | 2.74 | 3.528(3) | 149.2 |
| C(2B)-H(2B)···I(6)#3 | 0.95 | 3.12 | 3.853(3) | 135.5 |
| C(2A)-H(2A)···I(2)#2 | 0.95 | 3.08 | 4.002(3) | 164.2 |
| C(5B)-H(5B)···I(8)#5 | 0.95 | 3.22 | 4.090(3) | 152.3 |
| C(7B)-H(7B1)···I(2)#6 | 0.98 | 3.26 | 3.774(4) | 114.2 |
| C(7B)-H(7B2)···I(5)#7 | 0.98 | 3.33 | 3.966(4) | 124.7 |
| C(7B)-H(7B3)···I(7)#3 | 0.98 | 3.22 | 3.796(5) | 119.5 |
| C(5A)-H(5A)···I(6)#3 | 0.95 | 3.30 | 3.841(3) | 118.3 |

Symmetry transformations used to generate equivalent atoms:
#1 x,-y+1/2,z; #2 x-1/2,y,-z+3/2; #3 -x+1,-y+1,-z+1; #4 -x+1/2,-y+1,z-1/2; #5 x+1/2,y,-z+3/2; #6 -x+3/2,-y+1,z-1/2; #7 -x+3/2,-y+1,z+1/2

***Table SI5***. *Geometry of weak interactions in 4-CNpyBiI$_3$ [Å/°]*

| D-H···A | d(D-H) | d(H···A) | d(D···A) | <(DHA) |
|---|---|---|---|---|
| C(6A)-H(6A)...I(2)#2 | 0.93 | 3.33 | 4.170(8) | 151.7 |
| C(3A)-H(3A)...I(6) | 0.93 | 3.25 | 4.012(8) | 140.1 |
| C(2A)-H(2A)...I(4) | 0.93 | 3.23 | 4.100(8) | 157.0 |
| C(2A)-H(2A)...I(5) | 0.93 | 3.31 | 3.774(7) | 113.0 |
| O(1)-H(1)...I(7) | 0.82 | 2.91 | 3.597(10) | 142.8 |
| C(6B)-H(6B)...N(2B)#3 | 0.93 | 2.45 | 3.079(16) | 125.2 |
| N(1B)-H(1B)...O(1) | 0.86 | 1.85 | 2.631(15) | 149.5 |
| C(2B)-H(2B)...I(6)#4 | 0.93 | 3.14 | 3.966(13) | 149.2 |
| C(3B)-H(3B)...I(7)#4 | 0.93 | 3.13 | 4.058(13) | 175.4 |
| C(1)-H(1D)...I(1)#5 | 0.96 | 2.94 | 3.843(14) | 157.1 |
| C(7A)-H(7AB)...I(4) | 0.96 | 3.17 | 4.082(10) | 158.3 |

Symmetry transformations used to generate equivalent atoms:
#1 #2 x-1/2,-y+1/2,z-1/2 #3 -x+3/2,y+1/2,-z+1/2
#4 -x+2,-y,-z+1 #5 x+1,y,z

***Table SI6***. *Geometry of weak interactions in (3) [Å/°]*

| D-H···A | d(D-H) | d(H···A) | d(D···A) | <(DHA) |
|---|---|---|---|---|
| N(1A)-H(1A)···I(9)#1 | 0.88 | 2.70 | 3.560(10) | 166.0 |
| C(6A)-H(6A)···I(6)#2 | 0.95 | 3.25 | 4.085(11) | 147.9 |
| C(2C)-H(2C)···I(4)#3 | 0.95 | 3.24 | 3.991(12) | 137.2 |
| C(2C)-H(2C)···I(1)#3 | 0.95 | 3.22 | 3.710(11) | 114.0 |
| N(1B)-H(1B) ···I(5) | 0.88 | 2.75 | 3.604(11) | 163.6 |
| N(1C)-H(1C)···I(3) | 0.88 | 2.71 | 3.545(11) | 159.0 |
| C(2A)-H(2A)···I(9)#4 | 0.95 | 3.15 | 3.946(12) | 143.0 |
| C(3A)-H(3A)···I(5)#4 | 0.95 | 3.30 | 4.187(11) | 157.1 |
| C(6B)-H(6B)···I(4) | 0.95 | 3.07 | 3.797(13) | 134.0 |
| C(6C)-H(6C)···I(8)#5 | 0.95 | 3.22 | 4.118(13) | 158.9 |
| C(5A)-H(5A)···I(1)#2 | 0.95 | 3.14 | 4.051(12) | 160.3 |
| C(5B)-H(5B)···I(1)#2 | 0.95 | 3.03 | 3.791(11) | 138.7 |
| C(5B)-H(5B)···I(2)#6 | 0.95 | 3.27 | 3.941(12) | 129.2 |
| C(5C)-H(5C)···I(7)#7 | 0.95 | 3.13 | 4.073(13) | 170.1 |

Symmetry transformations used to generate equivalent atoms:
#1 x-1/2,-y+3/2,-z+1; #2 x,y+1,z; #3 x+1,y,z ; #4 x-1,y,z; #5 x,y-1,z; #6 -x,y+1/2,-z+1/2; #7 x+1/2,-y+1/2,-z+1

**(i)** **(ii)** **(iii)**

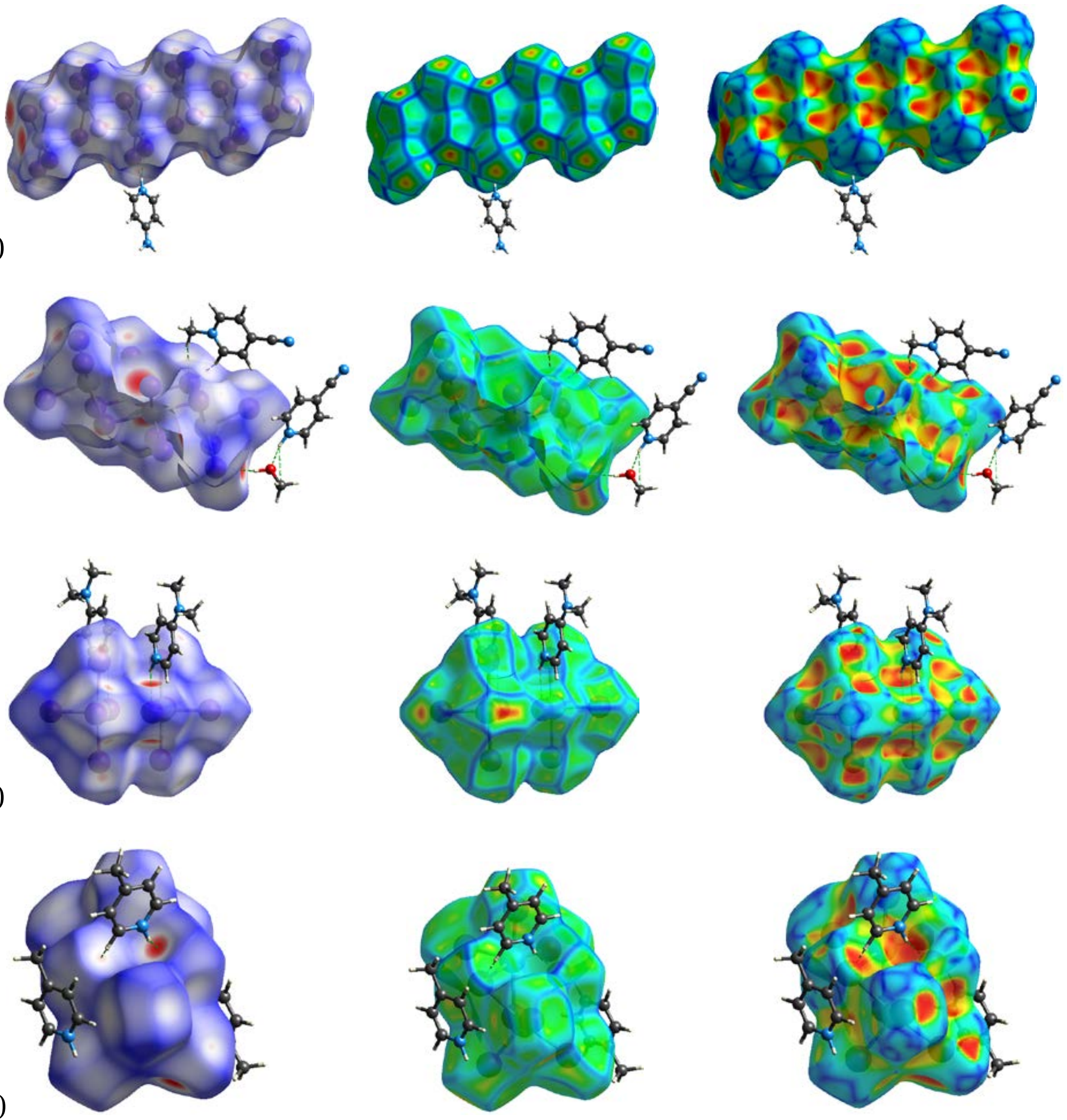

(a)

b)

(c)

(d)

*Figure SI 5.* ***(i)*** *3D $d_{norm}$ surface of (a) 4-Ampy$BiI_3$ in the range -0.4527 to 1.1099, (b) 4-Metpy$BiI_3$ ranging from -0.3428 to 1.2451 (c) 4-Dmetpy$BiI_3$ in the range -0.3012 to 1.1518, (d) 4-CNpy$BiI_3$ in the range of -0.3383 a.u. (red) to 1.3823a.u. (blue),* ***(ii)*** *curvedness in the range of (-4.0000 to 0.4000),* ***(iii)*** *shape index in the range of (-1.0000 to 1.0000).*

*(a)*

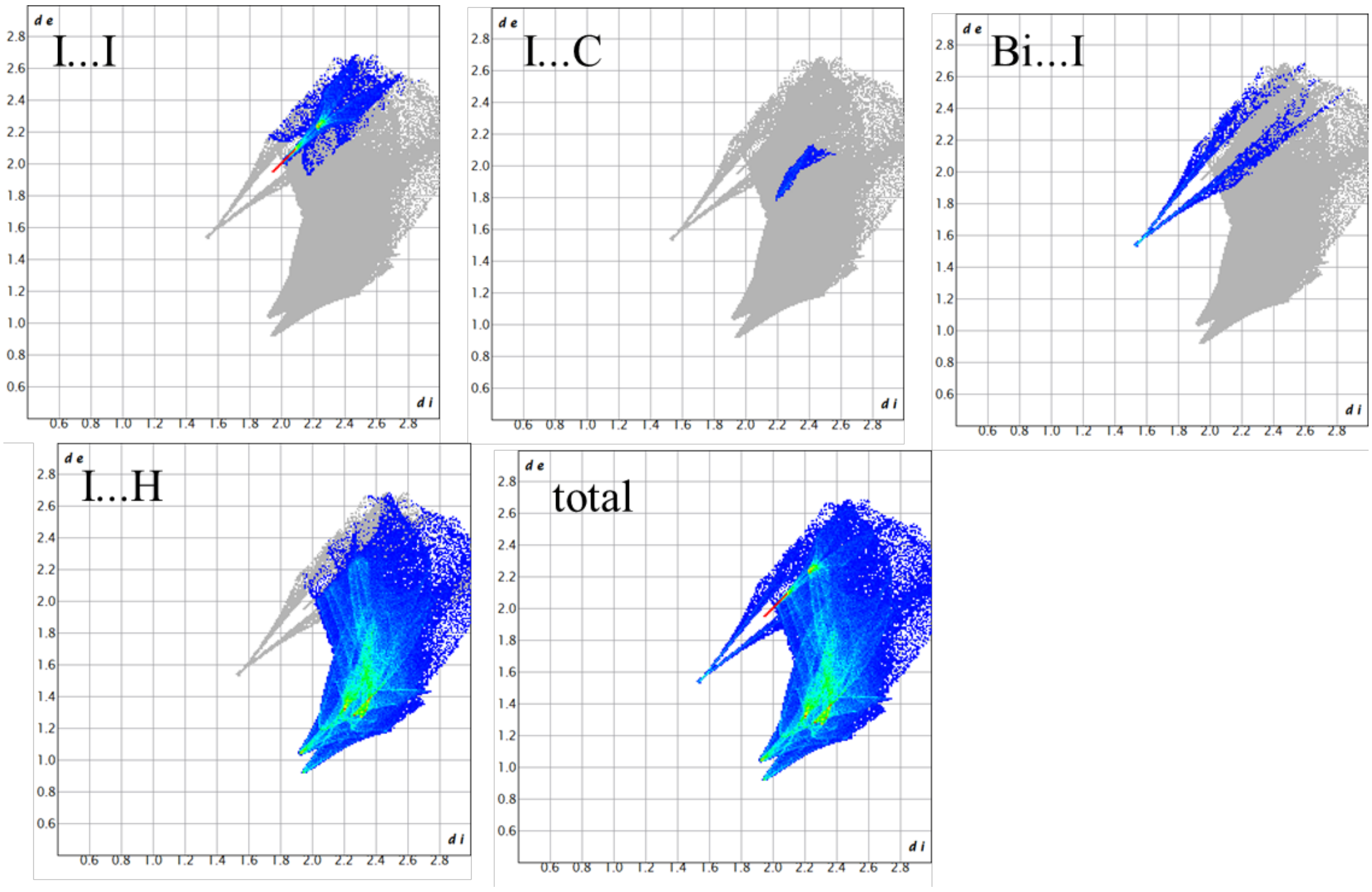

(b)

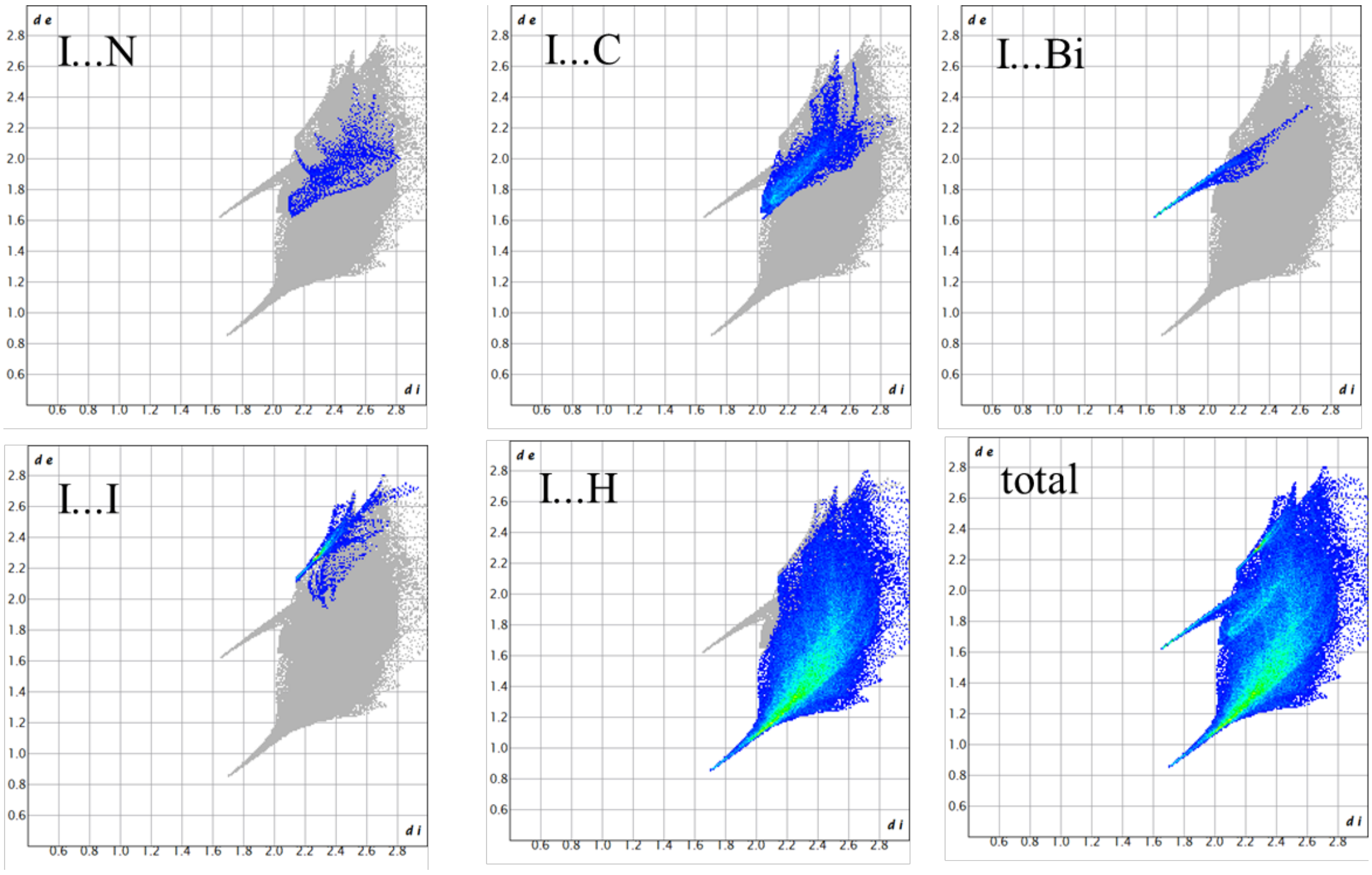

(c)

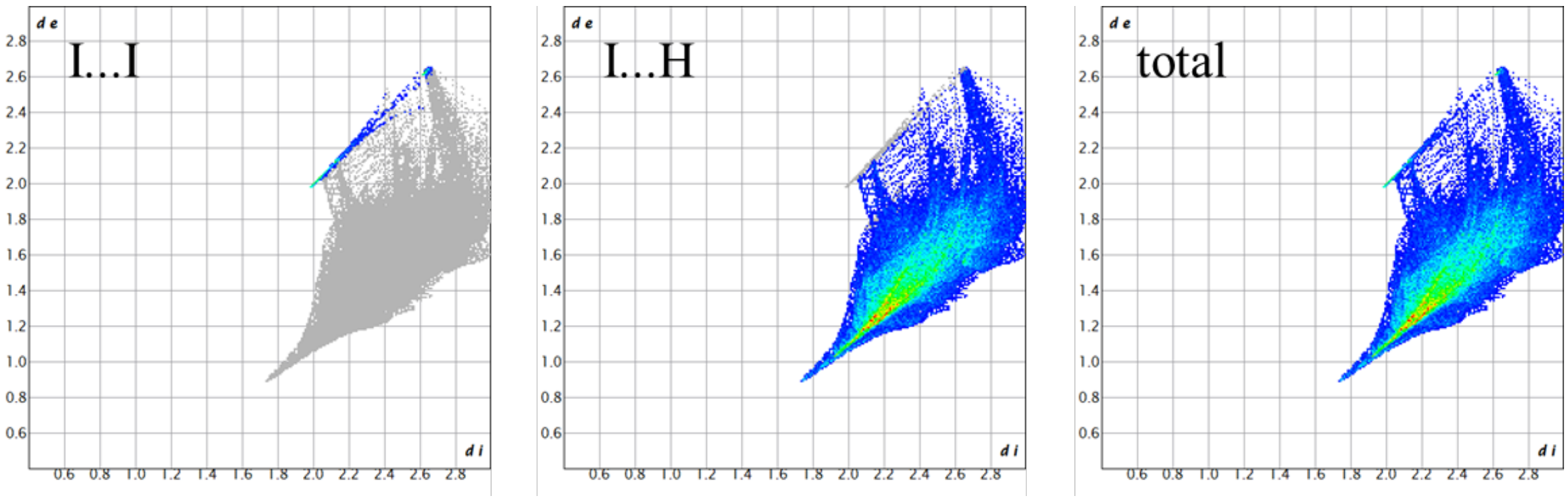

(d)

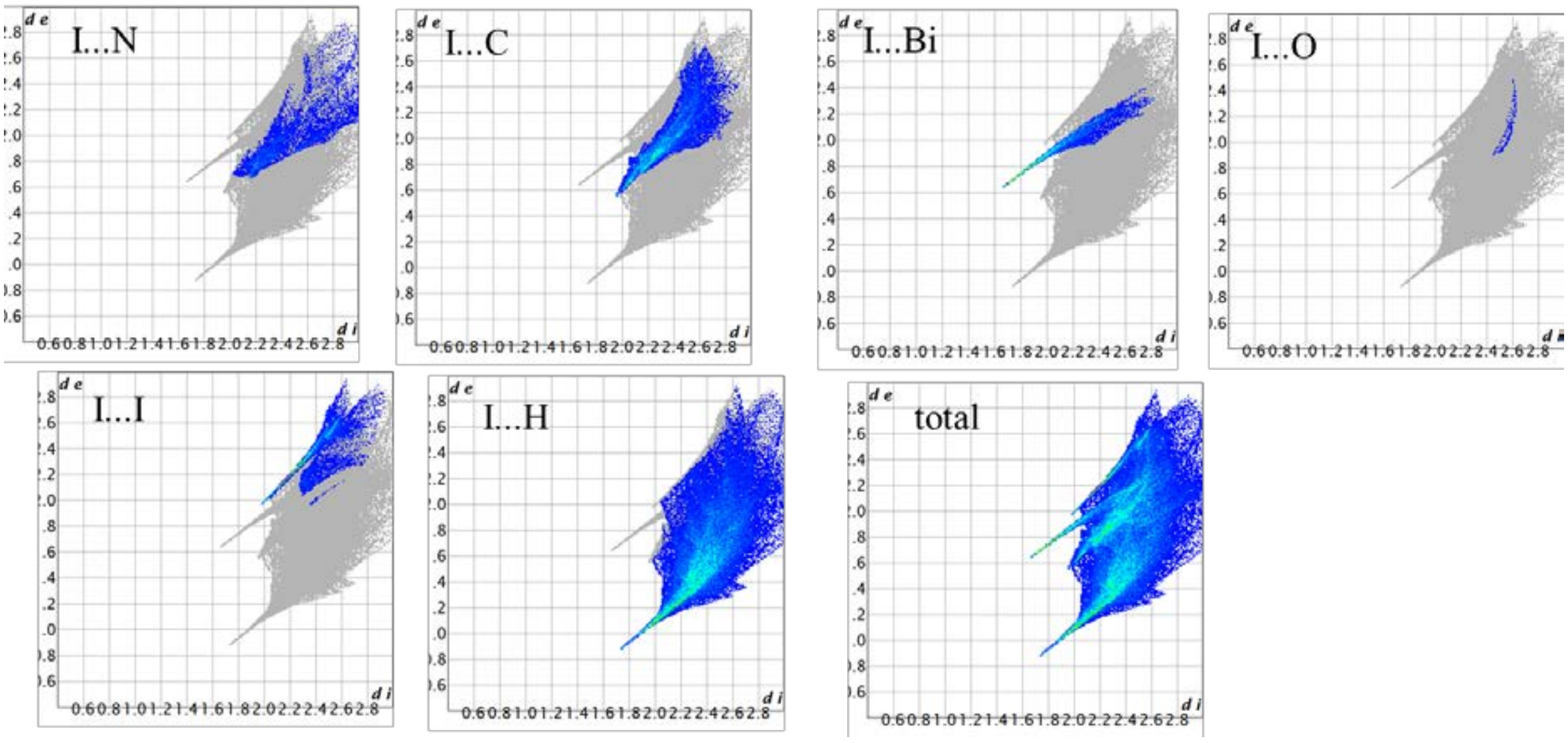

*Figure SI 6. Fingerprint plots of (a) 4-AmpyBiI$_3$, (b) 4-MetpyBiI$_3$ (c) 4-DmetpyBiI$_3$, (d) 4-CNpyBiI$_3$.*

***Table SI 7****. Percentage contributions of the various intermolecular contacts contributing to Hirshfeld's surfaces of the structures.*

| Percentage of interaction | 4-Ampy$BiI_3$ | 4-Mepy$BiI_3$ | 4-DMepy$BiI_3$ | 4-CNpy$BiI_3$ |
|---|---|---|---|---|
| I···I | 12.5 | 4.5 | 2 | 8.7 |
| I···H | 82.5 | 81 | 98 | 59.9 |
| I···N | 0.3 | 2 | - | 6.5 |
| I···C | 0.8 | 9.5 | - | 16.9 |
| Bi···I | 3.9 | 3 | - | 7.7 |
| Bi-O | - | - | - | 0.3 |

***Figure SI 7.*** *Minimum conduction band and maximum valence band structure and the dashed rectangular shows the region for calculation of the effective mass by the second derivative of the region.*

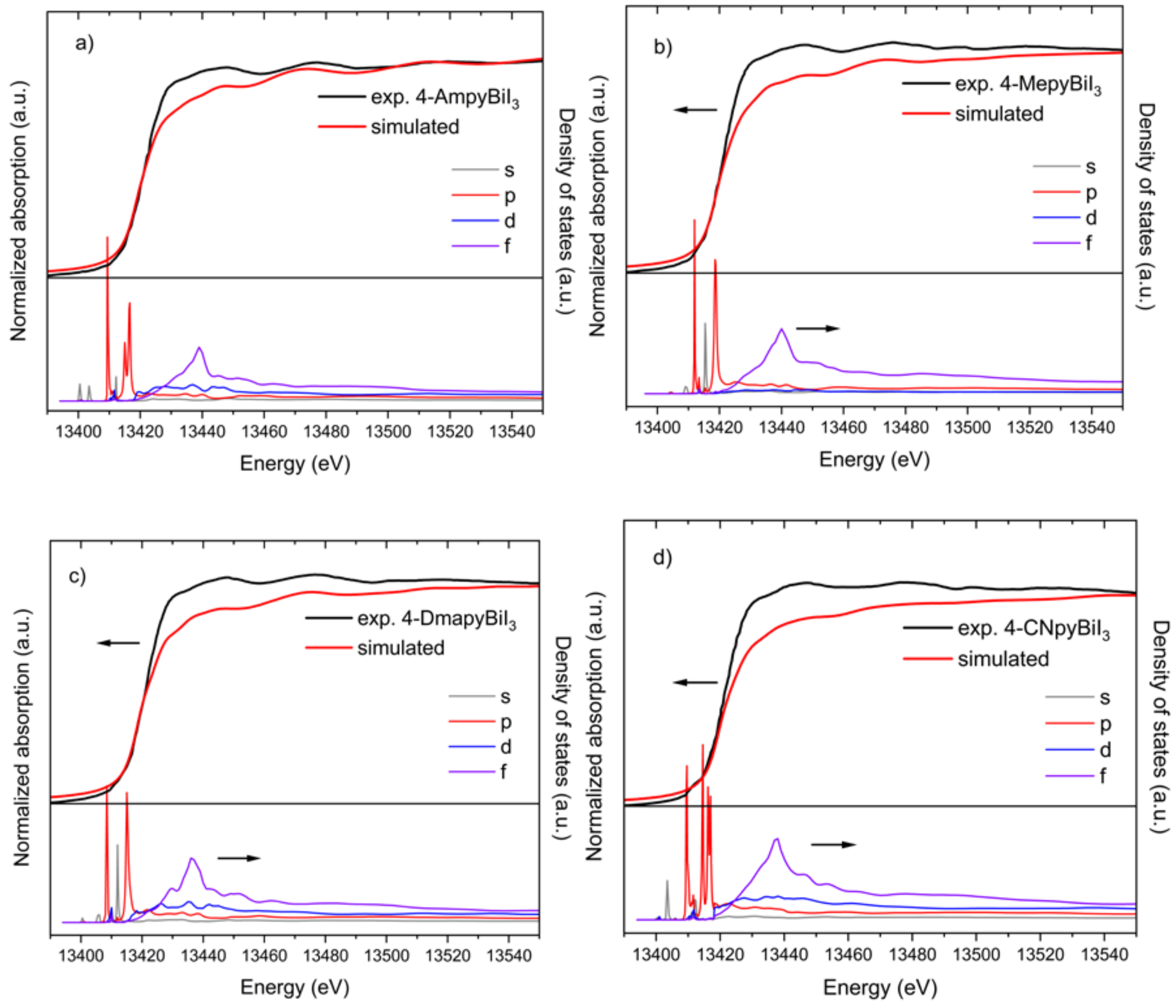

***Figure SI 8.*** *Experimental (black) and DFT-calculated (red) XAS spectra for 4-AmpyBiI*$_3$ *(a), 4-MepyBiI*$_3$ *(b), 4-DmapyBiI*$_3$ *(c) and 4-CNpyBiI*$_3$ *(d) at L3-edge are plotted on the left axis. The partial densities of the states projected on the absorbing Bi atom are also presented on the right axis.*

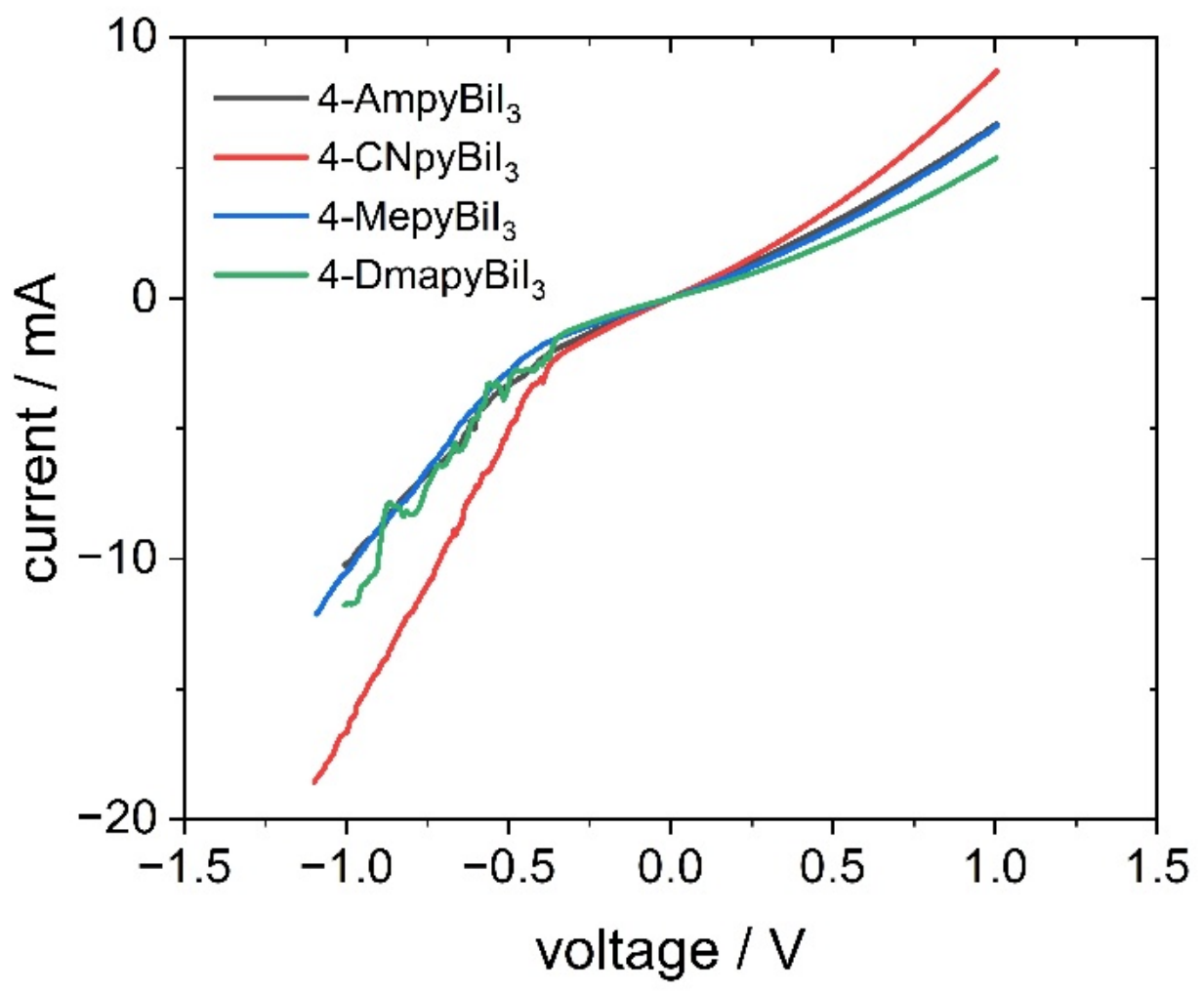

***Figure SI 9.*** *Current-voltage dependencies (linear scale) for all studied bismuth complexes recorded at 298 K.*